\documentclass{goose-article}

\hypersetup{pdfauthor={%
  T.W.J. de Geus%
}}
\header{%
  T.W.J.~de~Geus, R.H.J.~Peerlings, M.G.D.~Geers, 2016\\%
  International Journal of Solids and Structures, 97-98:687-698,  %
  \doi{10.1016/j.ijsolstr.2016.03.029}, \eprint{1603.05841}%
}

\title{%
  Competing damage mechanisms in a two-phase microstructure: how microstructure and loading conditions determine the onset of fracture%
}
\author[1,2]{T.W.J.~de~Geus$^*$}
\author[1]{R.H.J.~Peerlings}
\author[1]{M.G.D.~Geers}
\affil[1]{%
  Department of Mechanical Engineering, Eindhoven University of Technology, Eindhoven, The Netherlands%
}
\affil[2]{%
  Materials innovation institute (M2i), Delft, The Netherlands%
}
\contact{%
  $^*$Corresponding author:
  \href{mailto:t.w.j.d.geus@tue.nl}{t.w.j.d.geus@tue.nl} %
  \hspace{1mm}--\hspace{1mm} %
  \href{mailto:tom@geus.me}{tom@geus.me} %
  \hspace{1mm}--\hspace{1mm} %
  \href{http://www.geus.me}{www.geus.me}%
}

\begin{document}

\maketitle

\begin{abstract}
This paper studies the competition of fracture initiation in the ductile soft phase and in the comparatively brittle hard phase in the microstructure of a two-phase material. A simple microstructural model is used to predict macroscopic fracture initiation. The simplicity of the model ensures highly efficient computations, enabling an comprehensive study: a large range of hard phase volume fractions and yield stress ratios, for wide range of applied stress states. Each combination of these parameters is analyzed using a large set of (random) microstructures. It is observed that only one of the phases dominates macroscopic fracture initiation: at low stress triaxiality the soft phase is dominant, but above a critical triaxiality the hard phase takes over resulting in a strong decrease in ductility. This transition is strongly dependent on microstructural parameters. If the hard phase volume fraction is small, the fracture initiation is dominated by the soft phase even at high phase contrast. At higher hard phase volume fraction, the hard phase dominates already at low phase contrast. This simple model thereby reconciles experimental observations from the literature for a specific combination of parameters, which may have triggered contradictory statements in the past. A microscopic analysis reveals that the average phase distribution around fracture initiation sites is nearly the same for the two failure mechanisms. Along the tensile direction, regions of the hard phase are found directly next to the fracture initiation site. This `band' of hard phase is intersected through the fracture initiation site by `bands' of the soft phase aligned with shear. Clearly, the local mechanical incompatibility is dominant for the initiation of fracture, regardless whether fracture initiates in the soft or in the hard phase.
\end{abstract}


\section{Introduction}

\subsection{Objective}

This paper studies the competition between the different fracture initiation mechanisms in the microstructure of multi-phase alloys. The focus is on a class of materials that compromises between strength and ductility by employing a microstructure that consists of two or more phases/materials with distinct mechanical properties: a comparatively hard but brittle (reinforcement) phase and a comparatively soft ductile phase. Examples include metal matrix composites, e.g.\ silicon-carbide particles embedded in an aluminum matrix, and advanced high strength steels, such as dual-phase steel where the martensite phase reinforces the ferrite matrix. It has been shown, experimentally and numerically, that both phases contribute to fracture, and that their relative contribution is dependent on the microstructural morphology, the relative amount of the phases, the contrast of mechanical properties between them, and the applied stress state.

This paper studies the competition of the different microstructural damage mechanisms, and the consequences on the macroscopically observed fracture properties (i.e.\ strength and ductility), for these parameters (hard phase volume fraction, phase contrast, and stress state). This furthermore enables the characterization of the quintessential phase distribution around fracture initiation sites. The results are compared to observations from the literature, for a specific combination of parameters.

A computational multi-scale approach is proposed which predicts macroscopic fracture initiation as a natural outcome of unit cell computations. The model is chosen quite general, relevant for a wide range of materials in which the considered micro-damage mechanism are present. It is composed of a microstructure with two phases: a (quasi-)brittle hard phase that is embedded in a ductile soft matrix. The hard phase thereby is able to accommodate a small amount of plastic deformation before fracture; significantly less than the soft phase. To include sufficient statistical fluctuations, the adopted multi-scale approach is constructed to be computationally inexpensive by incorporating only the essential micromechanics. It uses simple indicators for both fracture initiation mechanisms: for the hard phase the Rankine model for cleavage \cite{Argon1975, Beremin1981, Bao1991} and for the soft phase the Johnson-Cook model for ductile damage \cite{Johnson1985}. The microstructure is represented by an ensemble of two-dimensional periodic volume elements that each consist of equi-sized square cells in which the phases are randomly distributed. The ensemble, comprising many random microstructures, is assumed macroscopically representative in an average sense. The adopted simplifications with respect to reality enable a systematic study as: (1) the computations are fast (few finite elements are needed for accurate discretization) and thus allow the comparison of many different microstructures for large range of parameter variations, (2) variations in terms of composition (i.e.\ volume fraction) are well controlled, and (3) the identification of the average distribution of phases around fracture initiation sites is well defined and transparent.

\subsection{State of the art}

The macroscopic stiffness and yielding of the considered class of multi-phase materials as a function of the constituents is reasonably well understood and can be predicted using a variety of models, ranging from a simple rule of mixtures to involved multi-scale computations \cite{Choi2009, Sun2009, Deng2006, Heinrich2012, Povirk1995, Scheunemann2013}. Most of these models however provide limited accuracy or insight when it comes to failure. The state of the art for metal matrix composites and dual-phase steel can be found in several review papers \cite{Mortensen2010,Rashid1981,Tasan2014b}.

Early papers already recognized the relationship between strength and ductility on the macro-scale and the volume fraction and mechanical properties of the reinforcement phase on the micro-scale \cite{Davies1978a, Lawson1981, Speich1979}. It is well-known from many experimental and numerical studies that an increase in reinforcement volume fraction elevates the fracture strength but at the same time lowers the ductility \cite{Ahmad2000, Llorca1991, LeRoy1981}. Depending on the reinforcement volume fraction and its mechanical properties, the dominant fracture mechanism changes from ductile to brittle. Mummery and Derby \cite{Mummery1991} have observed that fracture initiation tends towards failure of the silicon-carbide reinforcement particles when the volume fraction or the size of the particles is increased. Lee et al.\ \cite{Lee2004} have performed different heat treatments on dual-phase steel, and found that specimens with a high martensite hardness (tempered at a low temperature) reveal cleavage fracture, while specimens with low martensite hardness (tempered at higher temperatures) evidence ductile fracture; see also \cite{Kang2007}. Using X-ray micro-tomography, Maire et al.\ \cite{Maire2007} found that fracture initiates at or near the particle--matrix interface for a comparatively soft matrix while it initiates by cracking of the particles for a comparatively hard matrix. In other multi-phase materials, such as dual-phase steel, these observations are not as extensive due to the chemical and crystallographic similarities of the phases. A reduction of fracture strain is finally observed with increasing triaxiality, whereby the decrease is stronger than in other ductile materials as brittle fracture is often observed for higher triaxialities. To complicate matters, inspection of the fracture surfaces of these so-called brittle fracture modes still reveals dimples \cite{Cox1974, Papaefthymiou2006, Uthaisangsuk2008,Prahl2007,Anderson2014}. Focusing on the role of the morphology of the microstructure, it has been observed that fracture is promoted if the reinforcement particles are clustered \cite{Lewandowski1989, Kim1981, Kim2000}. For dual-phase steel, voids are frequently observed near the harder martensite bands caused by the rolling during the production of commercial grades \cite{Avramovic-Cingara2009, Kadkhodapour2011, Tasan2010}.

To study the effect of morphology numerically, the model should include (part of) the morphological complexity of the material. Researchers frequently use a microstructure that is directly obtained by microscopy. However, to limit the high complexity and computational cost, the mechanical degradation that results in fracture is often omitted and replaced by a highly simplified criterion whereby fracture develops along the localization of plastic strain \cite{Choi2009, Sun2009, Kumar2006, Povirk1995}. To account for the effect of local degradation due to void nucleation or growth, Prahl and co-workers apply the Gurson--Tvergaard--Needleman (GTN) model for the ductile soft phase and the cohesive zone model for the hard phase to predict macroscopic stress--strain curves up to fracture with adequate agreement with experiments \cite{Prahl2007, Uthaisangsuk2008}. However, focus is thereby not put on the individual contributions of the two fracture mechanisms in relation to the microstructure. Vajragupta et al.\ \cite{Vajragupta2012} apply similar models to a microscopic analysis. They observe that fracture initiates in a narrow hard region and it then propagates through the surrounding soft phase. This approach is restricted through its strong dependence on the finite element discretization.

The studies above do not systematically relate the microstructure to the (initiation of) fracture in a statistical sense. Mostly because the studied systems are not large enough or the models too complex to include statistical variations, but also because systematically generating different morphologies is highly non-trivial. To overcome this Kumar et al.\ \cite{Kumar2006} have tried to find the critical morphological feature in which the damage is high, regardless of the morphology at further distance. They thereby considered several artificial microstructures which carried the geometrical statistics of the real microstructure. More recently, De Geus et al.\ \cite{DeGeus2015a} considered a large ensemble of random microstructures to calculate the average morphology around the initiation of fracture in the soft phase. It was found that a single grain of the soft phase with neighboring regions of the hard phase on both sides along the tensile direction and interrupted by regions of the soft phase in the shear directions, correlates to high damage levels. These observations are supported by experimental studies from the literature. Avramovic-Cingara et al.\ \cite{Avramovic-Cingara2009a} identified that fracture initiates at interfaces perpendicular to the tensile axis. Such an interface appears even more critical when two regions of the reinforcement phase are closely separated \cite{Kumar2006, Kadkhodapour2011, Kang2007, Segurado2003, Williams2010, Williams2012}. Most notably, Segurado and LLorca \cite{Segurado2006} considered a three-dimensional periodic volume element containing spherical reinforcement particles. It was found that damage was preferentially nucleated in or in-between particles that are closely separated along the tensile axis. In a more pronounced form, it was found that a band of hard phase interrupted by the soft phase is critical for damage \cite{Avramovic-Cingara2009, Kadkhodapour2011, Tasan2010}.

\subsection{Outline}

The paper is structured as follows. The micromechanical model including the numerical implementation is discussed in Section~\ref{sec:model}. A reference ensemble of microstructures with a certain hard phase volume fraction and phase contrast is examined for a particular stress state in Section~\ref{sec:reference}. Using that ensemble, Section~\ref{sec:eta} studies the influence of different stress states. The hard phase volume fraction and the phase contrast are varied in Sections~\ref{sec:phi} and \ref{sec:chi}. All variations are combined to form a coherent mechanism map in Section~\ref{sec:mechanism}. The spatial distribution of phases around the fracture initiation sites is quantified in Section~\ref{sec:hotspot}, again for the reference ensemble. In Section~\ref{sec:ductile} the damage model for the hard phase is replaced by a -- completely different -- ductile criterion to assess the sensitivity of the results to the specific choice of damage model (for the reference ensemble). The paper ends with concluding remarks in Section~\ref{sec:conclusion}.

\section{Modeling}
\label{sec:model}

\subsection{Microstructure}

To obtain a statistically meaningful representation of the microstructure, an ensemble of $256$ random volume elements is used. Although the analysis is done on the ensemble as a whole, the mechanical response is computed on each of the volume elements separately, whereby periodicity eliminates boundary effects. The resulting computation is considerably more efficient compared to a single, large, volume element. Also, the computation is naturally parallelized.

\begin{figure}[htp]
  \centering
  \includegraphics[width=0.7\linewidth]{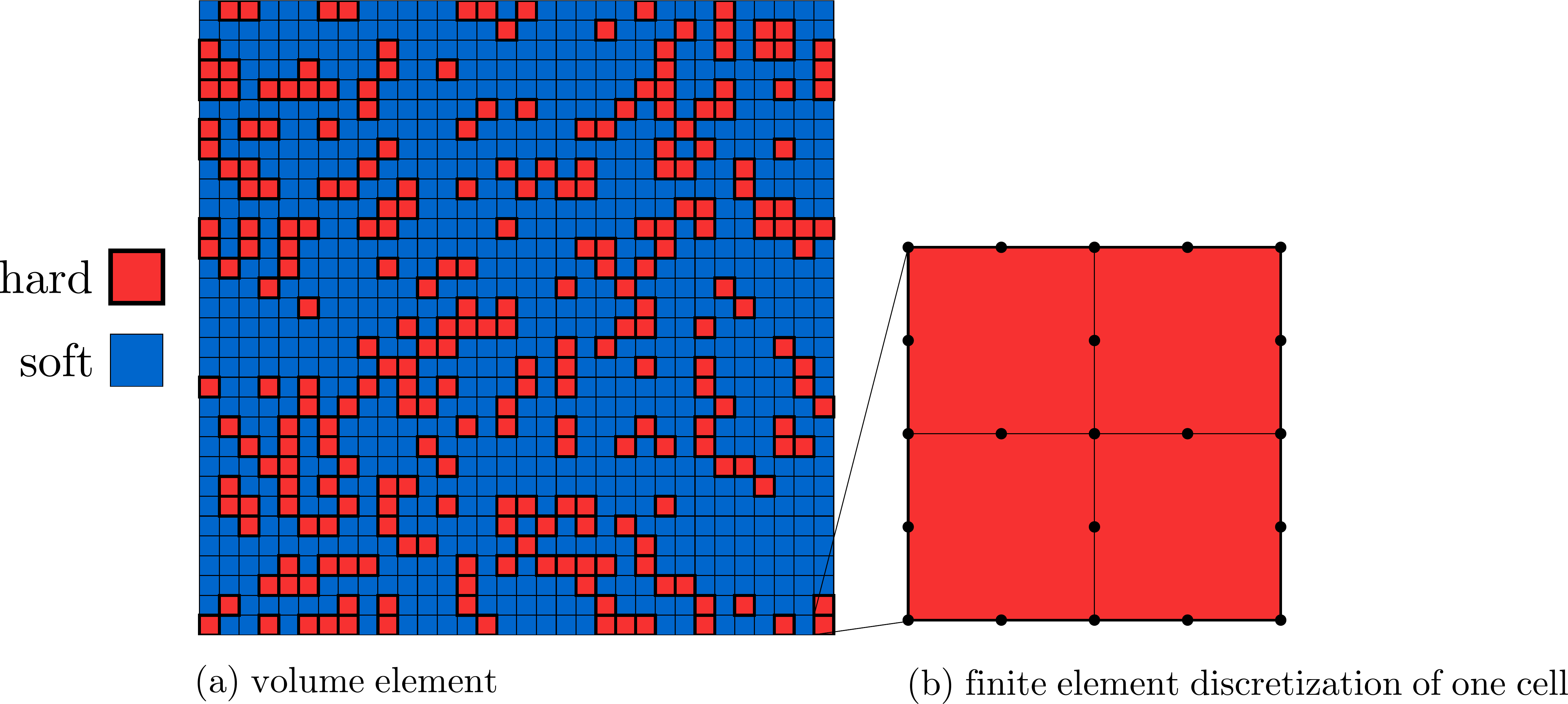}
  \caption{(a) A typical volume element from the ensemble with $\varphi^\mathrm{hard} = 0.25$. The hard phase is shown in red and the soft phase in blue. (b) The finite element discretization of a cell (all cells are discretized in the same way); the nodes are shown using black markers.}
  \label{fig:typical_geometry}
\end{figure}

Each of the periodic volume elements comprises $32 \times 32$ square cells that represent the individual grains or particles of the material. Each individual cell is randomly assigned the properties of either the soft phase or the hard phase by comparing a random number in the range $[0,1]$ to a probability $\varphi^\mathrm{hard}$. The resulting hard phase volume fraction is close to $\varphi^\mathrm{hard}$ and typically varies between $\pm 5\%$ of it. A reference case is chosen with $\varphi^\mathrm{hard} = 0.25$, for which a typical volume element is shown in Figure~\ref{fig:typical_geometry}(a) using red for the hard phase and blue for the soft phase, but $\varphi^\mathrm{hard}$ is also varied.

The response is calculated using the finite element method. Given the assumed idealization, only cell averaged quantities are considered. The finite element discretization is chosen such that the averaged quantities are independent of it. Each cell is discretized using $2 \times 2$ eight-node quadratic quadrilateral finite elements. Numerical integration is performed using four Gauss-points per finite element. The considered tensor components and scalar quantities are volume averaged over all $16$ Gauss-points in the cell. It has been verified that this discretization is sufficiently accurate, whereby it was found that the local relative error is 1\% with respect to a reference discretization of $10 \times 10$ finite elements per cell in terms of the stress and (plastic) strain components.

\subsection{Constitutive model}

Both phases are assumed isotropic elasto-plastic. At the moment of the initiation of fracture, the local deformations are large, and the response can be well in the plastic regime. A constitutive model suitable for such conditions is the model due to Simo \cite{Simo1992a}. This model uses a linear relation between the Kirchhoff stress $\bm{\tau}$ and the logarithmic elastic strain $\tfrac{1}{2} \ln (\bm{b}_\mathrm{e})$ (where $\bm{b}_\mathrm{e}$ is the elastic Finger tensor), involving the conventional elasticity tensor with the Young's modulus $E$ and Poisson's ratio $\nu$.

The plasticity is modeled using $J_2$-plasticity. In accordance with this model an associative flow rule is used. Linear hardening is assumed, which corresponds to the following yield function:
\begin{equation}
  \Phi( \bm{\tau}, \varepsilon_\mathrm{p} ) =
  \tau_\mathrm{eq} (\bm{\tau}) - (\tau_\mathrm{y0} + H \varepsilon_\mathrm{p})
  \leq 0
\end{equation}
where $\tau_\mathrm{eq}$ is the equivalent stress and $\varepsilon_\mathrm{p}$ is the accumulated equivalent plastic strain; the initial (Kirchhoff) yield stress $\tau_\mathrm{y0}$ and the hardening modulus $H$ are material parameters. The flow rule follows from normality.

The parameters of the soft phase are kept constant throughout this paper. They are chosen loosely representative for a specific class of materials, namely dual-phase steel \cite{Sun2009, Sun2009a, Al-Abbasi2003, Asgari2009, Vajragupta2012}, as follows:
\begin{equation}
  \frac{\tau_\mathrm{y0}^\mathrm{soft}}{E} = 3 \cdot 10^{-3} \qquad
  \frac{H^\mathrm{soft}}{E} = 4 \cdot 10^{-3} \qquad
  \nu = 0.3
\end{equation}
The hard phase differs from the soft phase only through the plastic response. The yield stress of the hard phase is related to that of the soft phase through the phase contrast factor $\chi$, i.e.\
\begin{equation}
  \tau_\mathrm{y}^\mathrm{hard}
  = \chi \; \tau_\mathrm{y}^\mathrm{soft}
  = \chi \left( \tau_\mathrm{y0}^\mathrm{soft} + H^\mathrm{soft} \varepsilon_\mathrm{p} \right)
\end{equation}
The value of $\chi = 2$ is used as a reference, but the influence of $\chi$ is also studied.

\subsection{Applied deformation and stress state}

The periodicity of the volume element is enforced by nodal tyings along the edge of the unit cell. Only the average displacement is prescribed while all fluctuations along the boundaries and throughout the volume element are permitted. As a reference load case, macroscopic pure shear deformation is prescribed in combination with a plane strain condition for the out-of-plane direction. This corresponds to the following macroscopic logarithmic strain tensor
\begin{equation}
\label{eq:model:pure-shear}
  \bar{\bm{\varepsilon}} = \bar{\bm{\varepsilon}}_\mathrm{d} =
  \frac{\sqrt{3}}{2} \, \bar{\varepsilon}\,
  \left(
    \vec{e}_\mathrm{x} \vec{e}_\mathrm{x} -
    \vec{e}_\mathrm{y} \vec{e}_\mathrm{y}
  \right)
\end{equation}
where $\bar{\varepsilon}$ is the macroscopic equivalent logarithmic strain. It is prescribed in small increments of $0.1 \%$ until macroscopic fracture initiation is predicted (discussed in the next section).

The different considered stress states are applied as a variation of the pure shear deformation. Since \eqref{eq:model:pure-shear} is volume preserving, the macroscopic hydrostatic stress $\bar{\tau}_\mathrm{m} = 0$ and therefore also the macroscopic stress triaxiality $\bar{\eta} = 0$. The latter is defined in terms of the macroscopic Kirchhoff stress as
\begin{equation}
  \bar{\eta} = \frac{\bar{\tau}_\mathrm{m}}{\bar{\tau}_\mathrm{eq}}
\end{equation}
Different triaxialities $\bar{\eta}$ are applied, each constant throughout the deformation history. For efficiency reasons this is done by superimposing a hydrostatic component to the local stress distribution $\bm{\tau}$ obtained from the pure shear simulation as follows:
\begin{equation}
\label{eq:model:triax}
  \bm{\tau}^\star(\vec{x},\bar{\varepsilon}) =
  \bar{\eta} \; \bar{\tau}_\mathrm{eq} (\bar{\varepsilon}) \; \bm{I} +
  \bm{\tau}(\vec{x},\bar{\varepsilon})
\end{equation}
Since the microstructure is elastically homogeneous, the resulting stress tensor $\bm{\tau}^\star$ is in equilibrium, however, the strain is no longer fully compatible with it, since the added hydrostatic stress would result in additional volumetric strain -- which is however elastic and hence small. It has been verified that the relative error in $\bar{\varepsilon}$ is less than 1\%, by comparing the approximated response to a full quasi three-dimensional computation subjected to a constant $\bar{\eta}$ \cite{DeGeus2015}. Note that in \cite{DeGeus2015}, the effect of the square cell shape was also found to be small when compared to hexagonal cells.

\subsection{Damage indicators}

Convincing experimental evidence revealed that in ductile materials, voids and/or cavities nucleate and grow throughout all stages of deformation. Then rapid and highly localized coalescence to global fracture occurs (\cite{LeRoy1981, Lewandowski1989, Avramovic-Cingara2009a, Kadkhodapour2011, Prahl2007} and many others). This behavior is modeled using damage descriptors that identify fracture initiation in the individual cells (representing grains or particles). These individual damage events are assumed not to interact strongly up to global fracture, and are therefore not coupled to the mechanical response. Global fracture is predicted when a critical number of cells in the ensemble of microstructures have `fractured'.

The hard phase fails through cleavage fracture, for which a stress based Rankine damage descriptor is used \cite{Argon1975, Beremin1981, Bao1991}. Fracture initiates when the maximum principal stress in a cell, $\tau_\mathrm{I}$, reaches a critical value, $\tau_\mathrm{c}$. The damage indicator is defined accordingly as
\begin{equation} \label{eq:model:D_hard}
  D = \frac{ \tau_\mathrm{I} }{ \tau_\mathrm{c} }
\end{equation}
so that $D = 0$ initially and fracture initiation is predicted for $D = 1$. Naturally, $\tau_\mathrm{c}$ is a material parameter (see below).

The soft phase is assumed to fail in a ductile manner, which is characterized by the Johnson-Cook model \cite{Johnson1985}. Reformulated in an incremental form, this model compares the rate of effective plastic strain in a cell, $\dot{\varepsilon}_\mathrm{p}$ (which is by definition non-negative), to a critical strain $\varepsilon_\mathrm{c}$ as follows
\begin{equation} \label{eq:model:D_soft}
  D =
  \int_0^t \frac{\dot{\varepsilon}_\mathrm{p}}{\varepsilon_\mathrm{c} (\eta)}
  \; \mathrm{d} \tau
\end{equation}
The critical strain $\varepsilon_\mathrm{c}$ depends on the stress triaxiality in that cell in the following way:
\begin{equation}
  \varepsilon_\mathrm{c} = A \exp \left( - B \eta \right) + \varepsilon_\mathrm{pc}
\end{equation}
where the parameters $A$, $B$, and the critical plastic strain $\varepsilon_\mathrm{pc}$ are material parameters.

The material parameters for both failure mechanisms are taken from the literature in the same range as the parameters of the constitutive model \cite{Vajragupta2012}:
\begin{equation} \label{eq:model:D_param}
  \frac{\tau_\mathrm{c}}{E} = 6.4 \cdot 10^{-3} \qquad
  A = 0.2 \qquad B = 1.7 \qquad
  \varepsilon_\mathrm{pc} = 0.1
\end{equation}
For the homogeneous hard phase, this implies that maximum 5\% plastic strain is allowed under uniaxial tension (see also \citep{Bareggi2012}).

Based on both damage indicators in Eqs.~(\ref{eq:model:D_hard},\ref{eq:model:D_soft}), an indicator for fracture initiation is defined on the cell level, denoted by $\mathcal{D}$: $\mathcal{D} = 1$ when fracture has initiated, which corresponds to any value $D \geq 1$, whereas otherwise $\mathcal{D} = 0$.

\subsection{Macroscopic fracture initiation}

Macroscopic fracture initiation follows in an averaged sense from local fracture initiation as described above. Macroscopic fracture is predicted when $1\%$ of the cells in the ensemble have `failed' -- i.e.\ $\mathcal{D} = 1$ in $1\%$ of the $256 \times 32 \times 32$ cells in the ensemble. Using the ensemble of different microstructures ensures that the predicted fracture initiation is representative and not due to a statistical fluctuation.

\section{Reference ensemble and load case}
\label{sec:reference}

\begin{figure}[htp]
  \centering
  \includegraphics[width=0.5\textwidth]{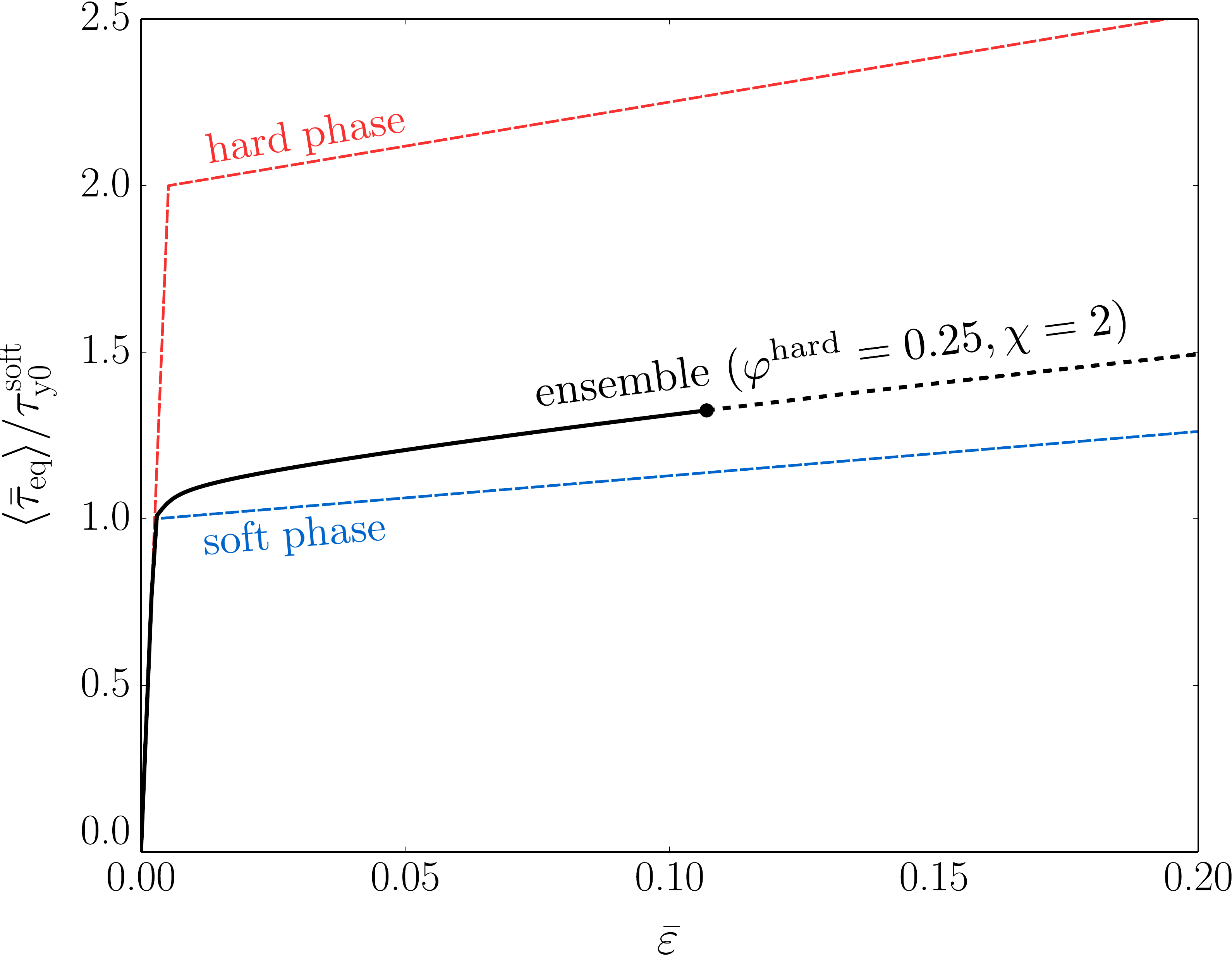}
  \caption{The macroscopic equivalent stress of the ensemble $\langle \bar{\tau}_\mathrm{eq} \rangle$ -- normalized by the initial yield stress of the ensemble $\tau_\mathrm{y0}^\mathrm{soft}$ -- as a function of the macroscopic, applied, equivalent strain $\bar{\varepsilon}$ for the reference configuration: $\varphi^\mathrm{hard} = 0.25$, $\chi = 2$, $\bar{\eta} = 0$ (black). The fracture initiation strain $\langle \bar{\varepsilon}_\mathrm{f} \rangle$ is indicated using a marker, the continued hardening response is shown with a dashed line. The constitutive response of the hard phase (red) and the soft phase (blue) are also included.}
  \label{fig:macroscopic_reference}
\end{figure}

The reference ensemble -- with hard phase volume fraction $\varphi^\mathrm{hard} = 0.25$, phase contrast $\chi = 2$, and applied triaxiality $\bar{\eta} = 0$ -- is considered first. The ensemble averaged macroscopic equivalent stress $\langle \bar{\tau}_\mathrm{eq} \rangle$ as a function of the applied macroscopic equivalent strain $\bar{\varepsilon}$ is shown using a solid black line in Figure~\ref{fig:macroscopic_reference}. For this curve, macroscopic fracture initiation is indicated using a marker and the hardening as predicted if loading is continued after fracture initiation is shown using a dashed black line. The constitutive response of the hard phase (red) and the soft phase (blue) are also included using dashed lines. As observed, macroscopic fracture initiation is predicted at a strain of $\langle \bar{\varepsilon}_\mathrm{f} \rangle = \bar{\varepsilon} = 0.11$\footnote{Although the strain is not averaged over the ensemble, as it is imposed on the individual volume elements, the notation $\langle ... \rangle$ is used also for the strain to emphasize that the fracture initiation criterion is defined on the ensemble as a whole.}. Compared to soft phase, the stress is increased by $16\%$ due to the introduction of the hard phase at the onset of macroscopic fracture. This increase comes at the expense of a decrease in fracture strain of $63\%$ compared to the homogeneous soft phase.

\begin{figure}[htp]
  \centering
  \includegraphics[width=1.0\linewidth]{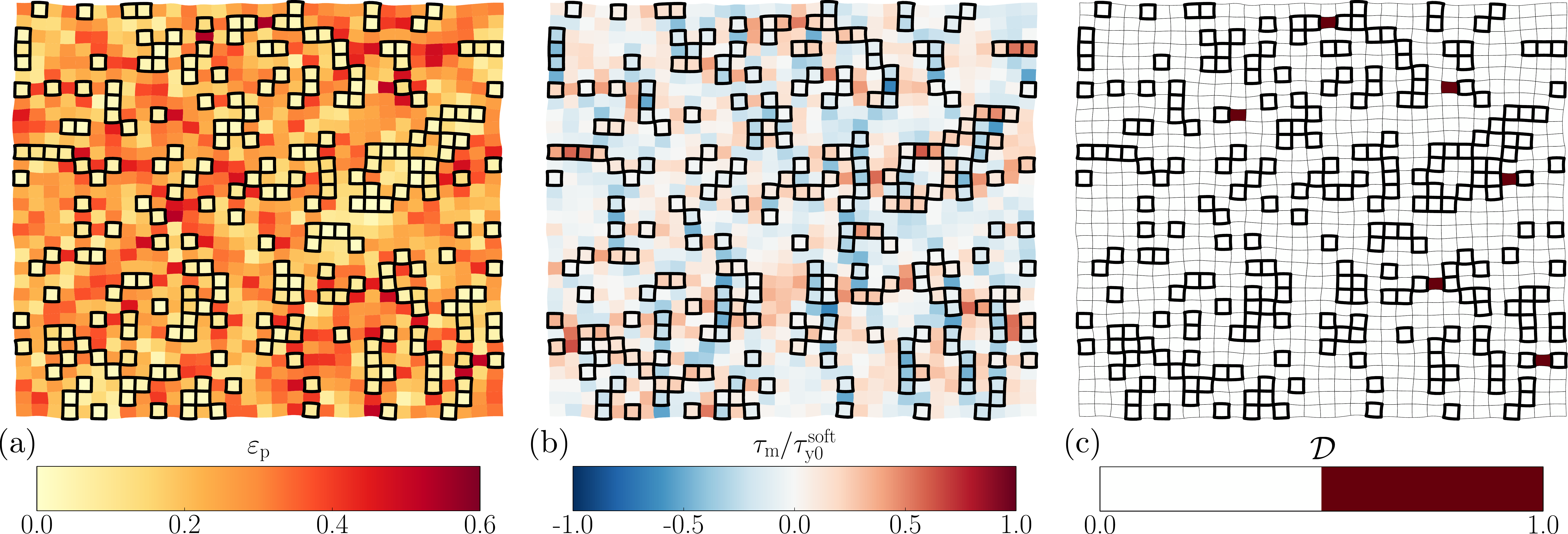}
  \caption{The response of the volume element of Figure~\ref{fig:typical_geometry} from the reference ensemble ($\varphi^\mathrm{hard} = 0.25$, $\chi = 2$, and $\bar{\eta} = 0$), at an applied macroscopic equivalent strain $\bar{\varepsilon} = \langle \bar{\varepsilon}_\mathrm{f} \rangle = 0.11$. From left to right: (a) the equivalent plastic strain $\varepsilon_\mathrm{p}$, (b) the hydrostatic stress $\tau_\mathrm{m}$ normalized by the initial yield stress of the soft phase $\tau_\mathrm{y0}^\mathrm{soft}$, and (c) the fracture initiation indicator $\mathcal{D}$. The hard phase cells are marked using a black outline.}
  \label{fig:typical_response}
\end{figure}

The microscopic response is visualized in Figure~\ref{fig:typical_response} for one microstructure from the ensemble, at the moment of macroscopic fracture initiation. The deformed geometry clearly shows the effect of the periodic boundary conditions, where the average deformation is prescribed by \eqref{eq:model:pure-shear} with extension in horizontal direction and compression in vertical direction. The individual cells are significantly deformed in all directions. The plastic strain, in Figure~\ref{fig:typical_response}(a), is largest in the soft phase, in particular where soft cells are interconnected under $\pm 45$ degree angles and surrounded by cells of the hard phase. This explains the decrease of macroscopic ductility observed above: the soft phase has to accommodate more (plastic) deformation due to phase contrast with the hard phase. The hydrostatic stress, in Figure~\ref{fig:typical_response}(b), has extremes in both phases, but is the largest in the hard phase. In the soft phase the highest tensile values are found where a soft cell is flanked by hard cells in the horizontal direction (e.g.\ on the bottom right). In the hard phase this is observed where several hard cells are linked in horizontal direction (e.g.\ on the top right). All predicted fracture initiation sites, in Figure~\ref{fig:typical_response}(c), are in the soft phase. In each case the soft cell is flanked by a hard cell on one or both sides.

\section{Influence of stress triaxiality}
\label{sec:eta}

\subsection{Results}

\begin{figure}[htp]
  \centering
  \includegraphics[width=0.5\textwidth]{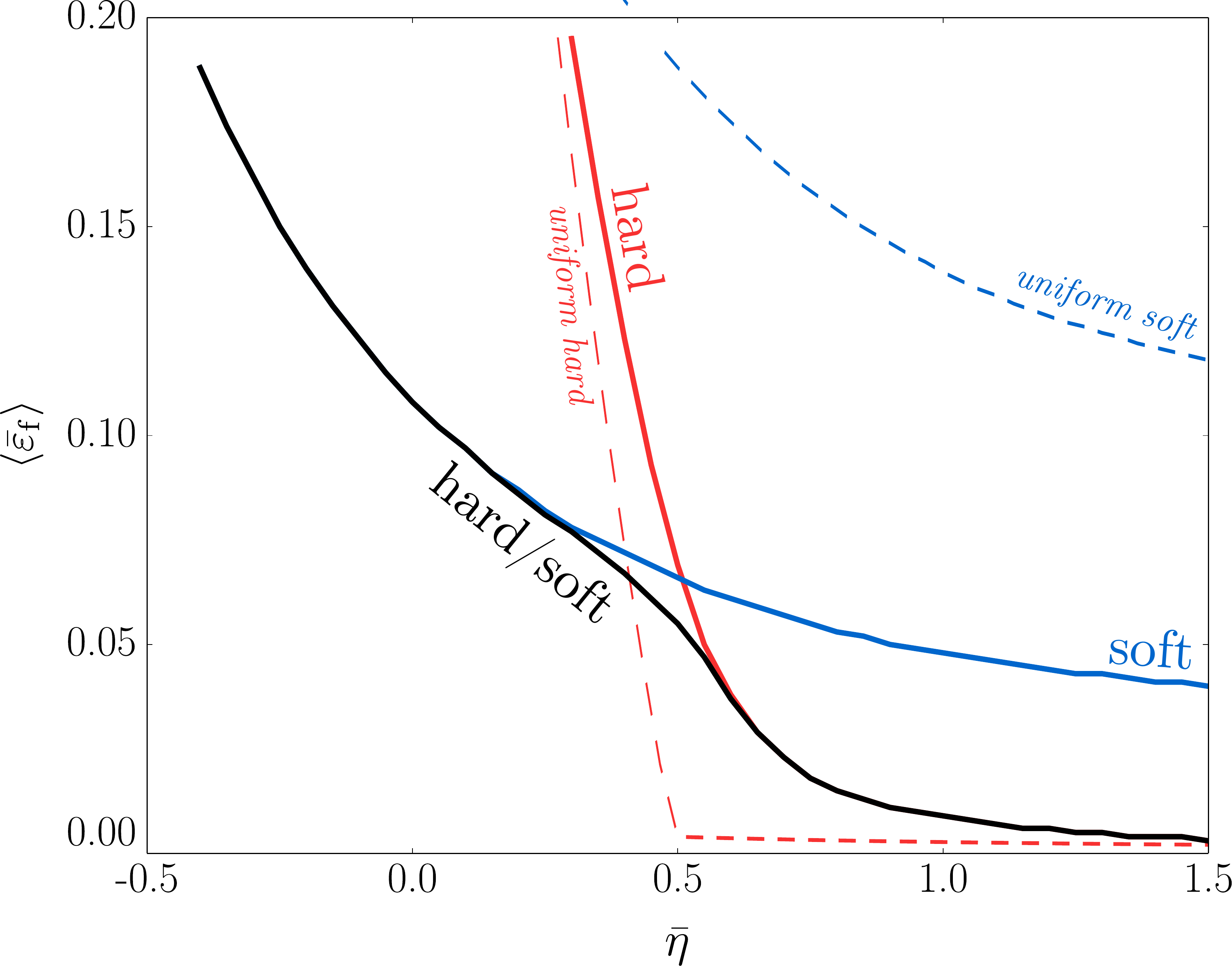}
  \caption{Macroscopic fracture initiation strain $\langle \bar{\varepsilon}_\mathrm{f} \rangle$ as a function of the applied triaxiality $\bar{\eta}$ for the reference configuration: $\varphi^\mathrm{hard} = 0.25$ and $\chi = 2$ (black). The fracture strains predicted by the isolated failure mechanisms are also shown: hard phase (red) and blue phase (soft) and for the uniform phases (dashed).}
  \label{fig:epsf_reference}
\end{figure}

The applied macroscopic stress triaxiality $\bar{\eta}$ is varied in the range $[-0.4,1.5]$. The resulting ensemble averaged fracture strain $\langle \bar{\varepsilon}_\mathrm{f} \rangle$ as a function of the applied triaxiality $\bar{\eta}$ is shown in black in Figure~\ref{fig:epsf_reference}, for the reference ensemble. Also shown are solid curves that correspond to the cases where failure is modeled only in the soft phase (i.e.\ fracture initiation is predicted in 1\% of all soft cells in the ensemble, in blue) or only in the hard phase (in red). The dashed curves correspond to the uniform soft phase (in blue) and hard phase (in red). Note that the number of `failed' cells varies between the different microstructures ranging from 2 to 21 (e.g.\ Figure~\ref{fig:typical_response}, wherein the number of `failed' cells equals 6).

Figure~\ref{fig:epsf_reference} shows that the fracture initiation strain $\langle \bar{\varepsilon}_\mathrm{f} \rangle$ is significantly reduced compared to a specimen of uniform soft phase. The overall trend is similar: the fracture initiation strain $\langle \bar{\varepsilon}_\mathrm{f} \rangle$ decreases for increasing triaxiality $\bar{\eta}$, as is frequently observed for ductile materials. However for the composite a rather rapid decrease of fracture strain is observed at a macroscopic triaxiality of $0.2 < \bar{\eta} < 0.5$. At low triaxialities, $\bar{\eta} < 0.2$, the macroscopic fracture initiation strain of the ensemble coincides with that of the soft phase (in blue), while at high triaxialities, $\bar{\eta} > 0.5$, it coincides with that of hard phase failure (in red).

The two failure mechanisms are thus in competition whereby the macroscopic triaxiality $\bar{\eta}$ plays a key role. At triaxialities $\bar{\eta} < 0.2$ a sufficient number of soft cells fail to call macroscopic fracture before a significant amount of damage is generated in the hard cells. For triaxialities $\bar{\eta} > 0.5$ the hard cells reach their failure criterion substantially earlier. In the range $0.2 < \bar{\eta} < 0.5$ the competition is more even and both contribute. Thus, even though fracture initiation occurs sooner at higher triaxialities for both damage mechanisms, the brittle failure mechanism becomes more pronounced due to its stress dependence that scales with the elastic modulus. The ductile failure mechanism on the other hand scales with the triaxiality through the plasticity via the much lower hardening modulus.

\begin{figure}[htp]
  \centering
  \includegraphics[width=1.0\linewidth]{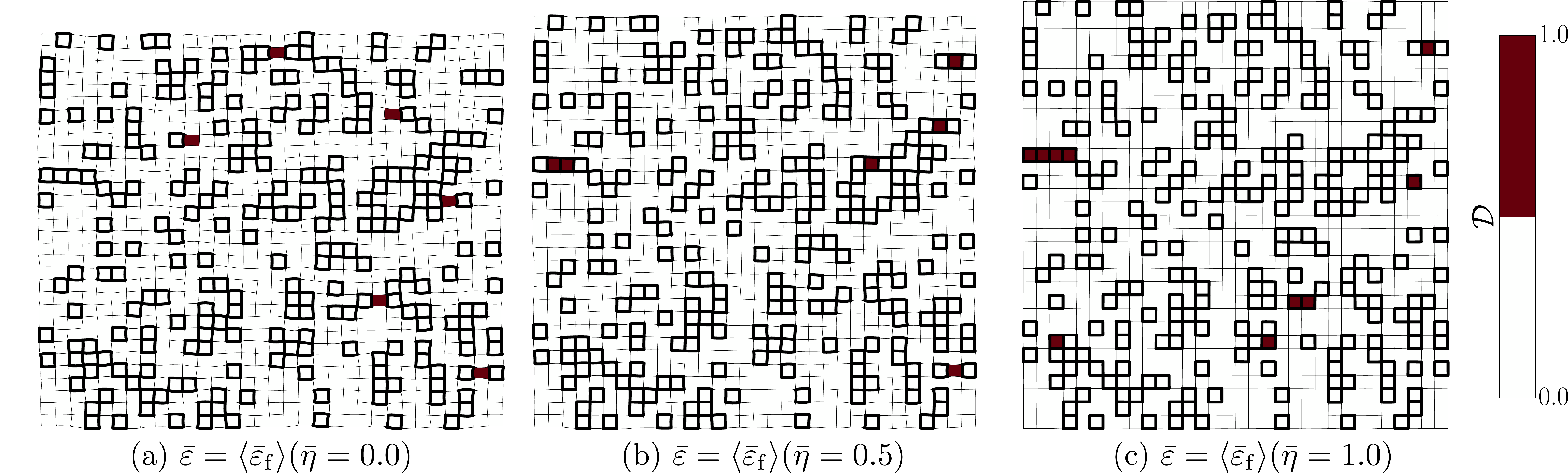}
  \caption{Fracture initiation in the microstructure of Figure~\ref{fig:typical_geometry} for different values of applied triaxiality. Note that the response is shown for different stages of deformation, corresponding to the respective macroscopic fracture initiation strain $\langle \bar{\varepsilon}_\mathrm{f} \rangle$ for each applied triaxiality $\bar{\eta}$ (see Figure~\ref{fig:epsf_reference}).}
  \label{fig:typical_damage}
\end{figure}

The trend in Figure~\ref{fig:epsf_reference} can be further analyzed by inspecting one typical microscopic response. The fracture initiation indicator $\mathcal{D}$ is shown in Figure~\ref{fig:typical_damage}, again for the volume element of Figure~\ref{fig:typical_geometry}. From left to right the triaxiality increases whereby the simulation is terminated at the relevant macroscopic fracture initiation strain $\langle \bar{\varepsilon}_\mathrm{f} \rangle$. For $\bar{\eta} = 0$ (Figure~\ref{fig:typical_damage}(a)) all fracture initiation sites are in the soft phase, always directly adjacent to hard phase. In the transition regime, for $\bar{\eta} = 0.5$ in Figure~\ref{fig:typical_damage}(b), several fracture initiation sites occur in hard cells. In the brittle fracture regime, for $\bar{\eta} = 1$ in Figure~\ref{fig:typical_damage}(c), all fracture initiation sites are in the hard phase. Like for the soft phase, each of these sites have hard phase to the left and/or right. The morphological characteristics of the phase distribution around the fracture initiation sites are discussed in more detail in Section~\ref{sec:hotspot}.

\subsection{Discussion}

The results indicate that the initiation of fracture is dominated by the ductile soft phase, by the brittle hard phase, or by both. The damage mechanisms in the two phases are thus in competition, whereby the outcome is determined by the macroscopic stress triaxiality, which is consistent with the literature \cite{Cox1974, Papaefthymiou2006, Uthaisangsuk2008, Prahl2007, Hoefnagels2015,Anderson2014,Paul2013a}. For example, \citet{Hoefnagels2015} carefully categorized and counted the damage events in a dual-phase steel subjected to different strain paths. In the context of the results above, their most important observation is that more fracture occurs in the hard martensite phase for a microstructure subjected to bi-axial loading (with a high triaxiality) compared with the uni-axial loading case (with a low triaxiality).

For multi-phase materials, few experimental studies have measured the macroscopic fracture strain as a function of stress triaxiality. A complicating factor is that different ranges of triaxialities require different sample geometries, associated with different macroscopic strain paths \citep{Bao2004a}. For sample geometries suitable for the shear regime, it has been observed that the fracture strain decreases with increasing triaxiality \citep{Anderson2014,Hauert2010,Lou2012a,Requena2013,Samei2016}, although the number of measurements are too few to be conclusive about the existence of a critical triaxiality upon which the ductility suddenly decreases. For sample geometries suitable for different strain paths, in particular for bi-axial tension, the fracture strain is observed to be higher than in pure shear \citep{Hauert2010,Lou2012a,Samei2016}. This observation invites further research as different strain paths have not been addressed in the present study, since they necessitate costly three-dimensional computations. Another question that is not addressed in the present work is whether different propagation mechanisms are triggered in different triaxiality regimes, as is known to be the case for more homogeneous materials \citep{Barsoum2007,Bao2004a}.

A strong limitation of the present study is the two-dimensional, plane strain, character of the microstructures. Recently similar computations have been performed on two- and three-dimensional microstructures \citep{DeGeus2016a}. The comparison showed that the damage distributions are similar for both cases, but the value of the damage is over-predicted using a two-dimensional model. For the present results this implies that the macroscopic fracture initiation strain, in Figure~\ref{fig:epsf_reference}, is lower than in a three-dimensional microstructure.

\begin{figure}[htp]
  \centering
  \includegraphics[width=.5\textwidth]{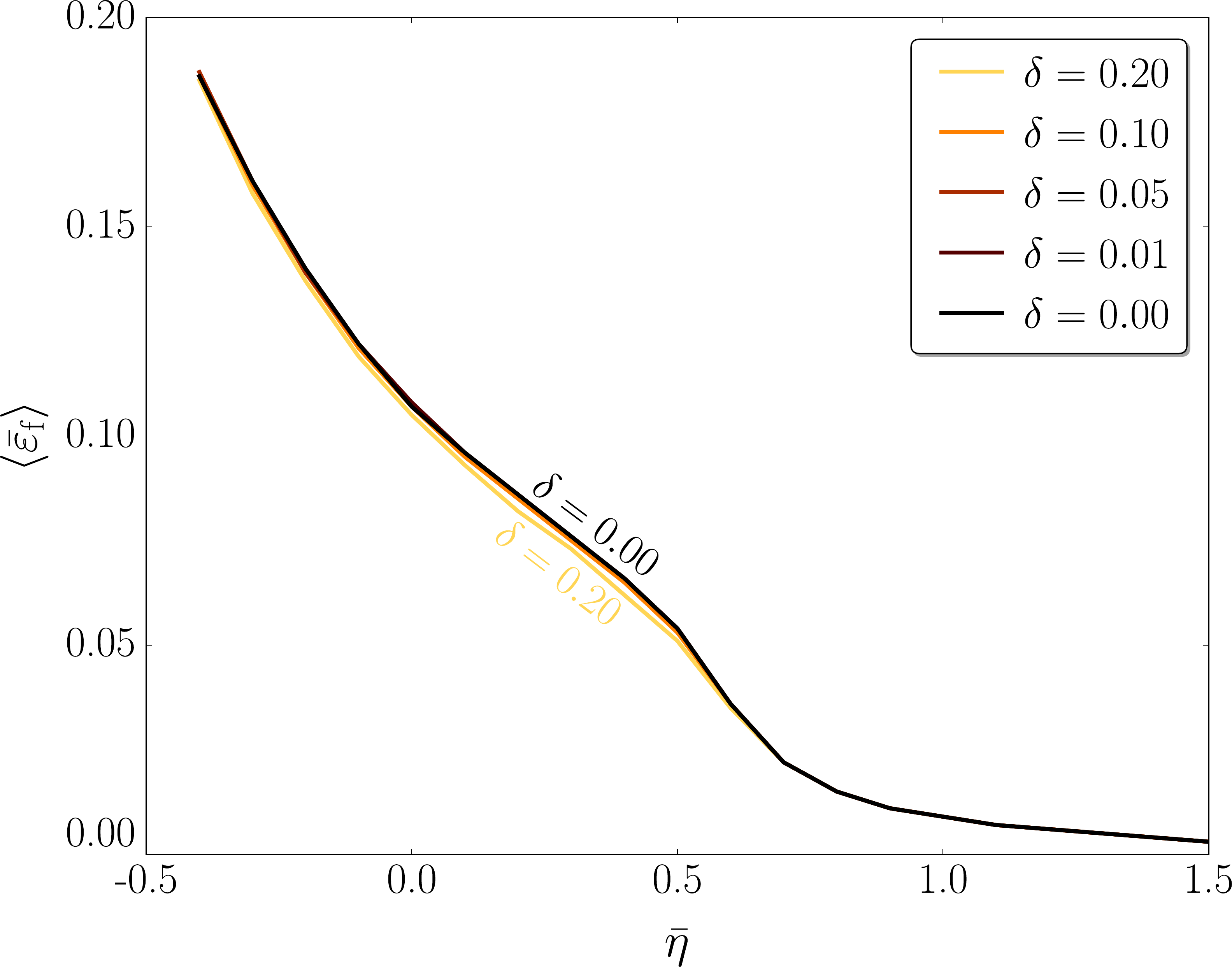}
  \caption{Effect of a random variation of the critical strain-to-failure between individual cells, with a standard deviation $\delta$. The reference $\delta = 0$ corresponds to uniform damage parameters throughout the microstructure.}
  \label{fig:epsf_influence-stochastics}
\end{figure}

For the present model, one may also ask to what extent the results depend on the definitions of microscopic and macroscopic fracture initiation. With respect to the former, in reality the strength may vary from cell to cell. This is modeled by randomly varying the strain-to-failure from cell to cell, for both phases. This critical strain is therefore multiplied with $1+\delta$ which randomly varies in spaces, with zero mean and a standard deviation for which the values $\delta = 0$, $0.01$, $0.05$, $0.1$, and $0.20$ are considered. The macroscopic fracture initiation strains as a function of the stress triaxiality are shown in Figure~\ref{fig:epsf_influence-stochastics}. Qualitatively they are unaffected by $\delta$, and also quantitatively the difference is small -- maximally a factor $50$ smaller than the applied variation. The reason of this is that the local phase distribution controls the damage, as is discussed in more detail in Section~\ref{sec:hotspot}.

\begin{figure}[htp]
  \centering
  \includegraphics[width=.5\textwidth]{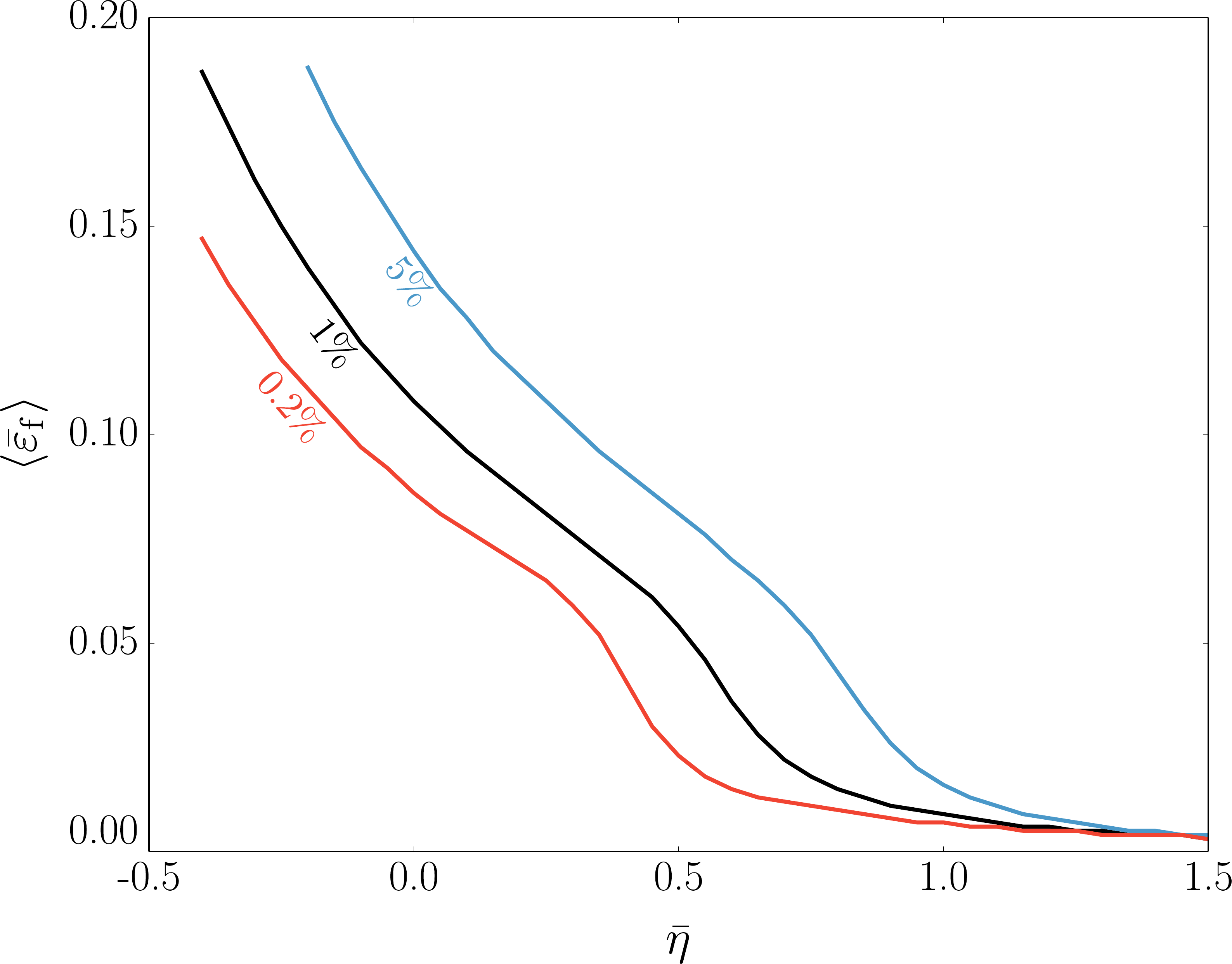}
  \caption{Effect of different macroscopic fracture criteria. Macroscopic fracture initiation is predicted respectively if $0.2\%$ (red), $1\%$ (black, reference), or $5\%$ (blue) of the cells in the ensemble have failed.}
  \label{fig:epsf_influence-n-crit}
\end{figure}

The definition for the macroscopic fracture initiation, i.e.\ that $1\%$ of the cells in the ensemble failed, is examined next. The analysis has therefore been repeated for the values $0.2$\% and $5\%$. The results, in Figure~\ref{fig:epsf_influence-n-crit}, show that the same qualitative trend is recovered in each case: ductile and brittle fracture are in competition. Quantitatively, the difference is $0.01$ in terms of fracture strain and approximately $0.18$ in terms of the critical stress triaxiality at which the failure switches from being dominated by the soft phase to being dominated by the hard phase. Future research that incorporates mechanical degradation would obsolete this simple criterion, by predicting macroscopic instability as a natural outcome of damage propagation on the micro-scale.

It is often hypothesized that since plasticity occurs in the hard phase, it fails through ductile fracture instead of brittle fracture \cite{Cai1985,Calcagnotto2010,Maresca2014,Steinbrunner1988,Ghadbeigi2010}. Therefore, the analysis is also carried out with a ductile fracture initiation criterion in the hard phase in Section~\ref{sec:ductile}.

\section{Influence of volume fraction}
\label{sec:phi}

In this section the effect of the variation of the hard phase volume fraction is considered. Ensembles with different hard phase volume fractions, $\varphi^\mathrm{hard}$, are used, yet each with the same reference phase contrast $\chi = 2$.

\begin{figure}[htp]
  \centering
  \includegraphics[width=.5\textwidth]{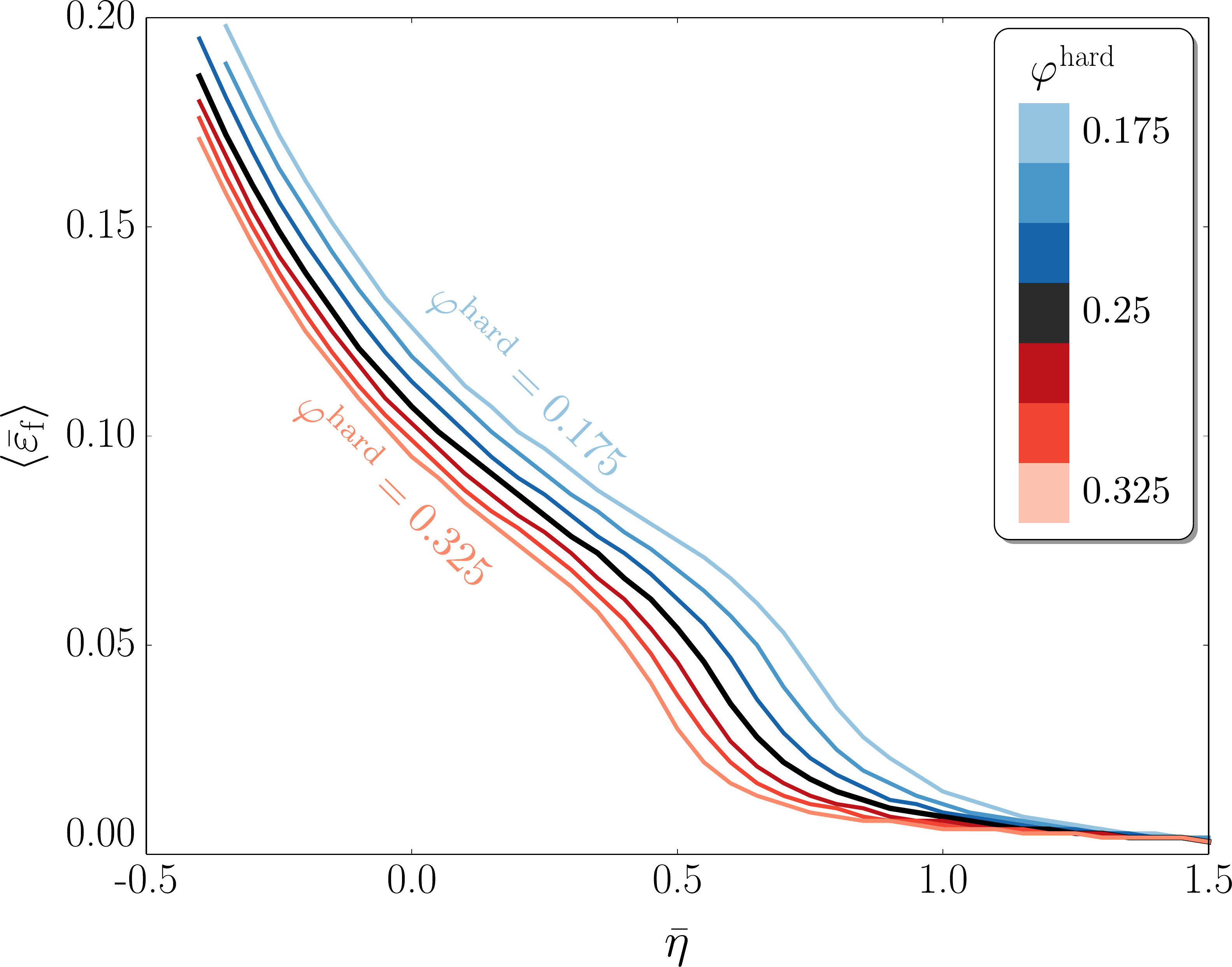}
  \caption{Macroscopic fracture initiation strain $\langle \bar{\varepsilon}_\mathrm{f} \rangle$ as a function of the applied triaxiality $\bar{\eta}$, for different hard phase volume fractions, $\varphi^\mathrm{hard}$, at constant phase contrast ($\chi = 2$). The colors indicate the different $\varphi^\mathrm{hard}$; the reference ensemble ($\varphi^\mathrm{hard} = 0.25$ and $\chi = 2$) in the middle is shown in black.}
  \label{fig:epsf_parameters_phivar}
\end{figure}

The macroscopic fracture initiation strain $\langle \bar{\varepsilon}_\mathrm{f} \rangle$ as a function of macroscopic stress triaxiality $\bar{\eta}$ is shown in Figure~\ref{fig:epsf_parameters_phivar} for seven different values of $\varphi^\mathrm{hard}$, each indicated using a different color. The reference ensemble of Figure~\ref{fig:epsf_reference}, is shown in black. Increasing the amount of hard phase promotes both the soft phase and the hard phase failure mechanisms. This is observed as a decrease of $\langle \bar{\varepsilon}_\mathrm{f} \rangle$ in both regimes. The triaxiality $\bar{\eta}$ at which the transition between the two regimes takes place decreases with increasing hard phase volume fraction.

\begin{figure}[htp]
  \centering
  \includegraphics[width=.5\linewidth]{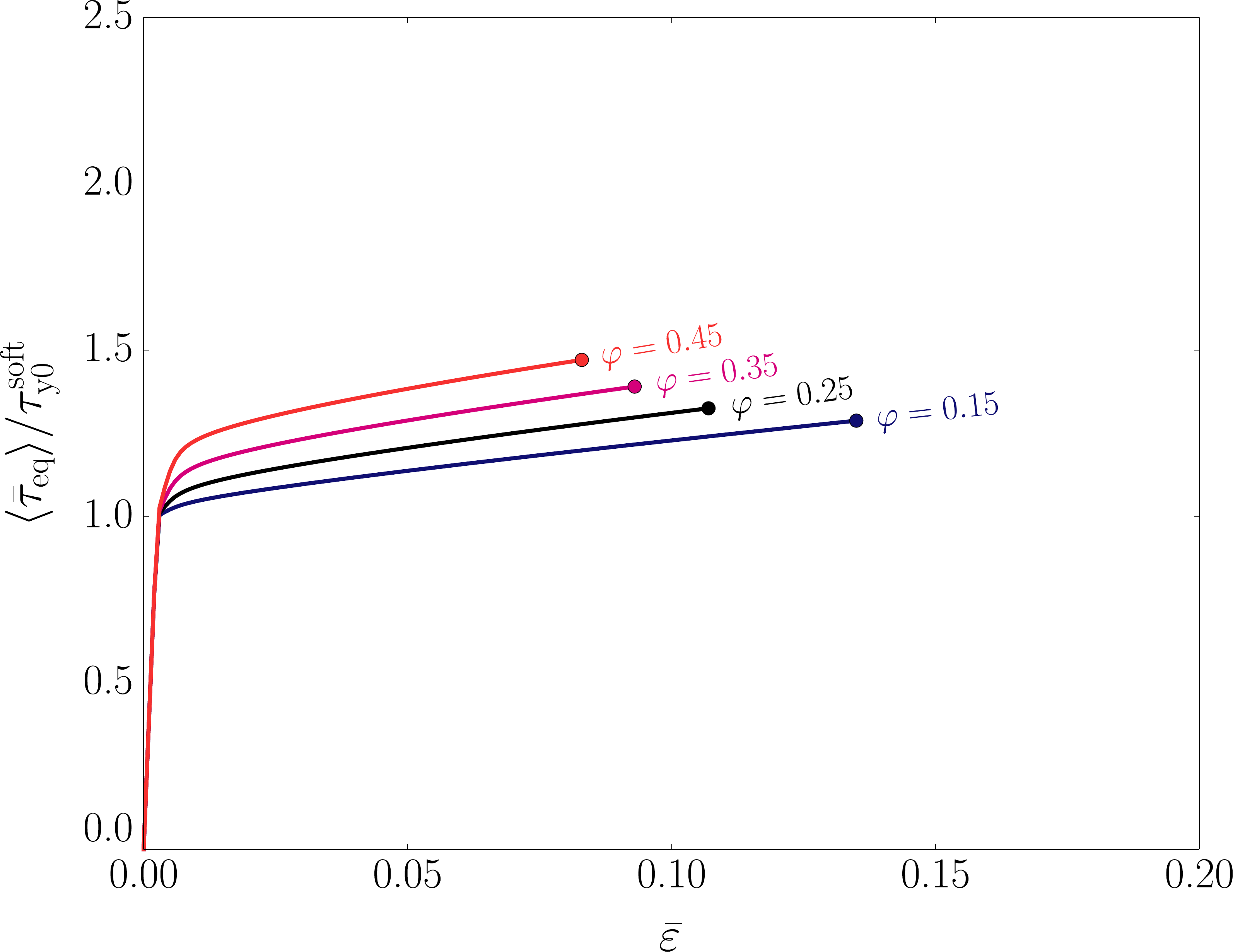}
  \caption{Ensemble averaged macroscopic equivalent stress $\langle \bar{\tau}_\mathrm{eq} \rangle$ as a function the macroscopic, applied, strain $\bar{\varepsilon}$ for different hard phase volume fractions $\varphi^\mathrm{hard}$ at constant phase contrast $\chi = 2$; all a constant applied triaxiality $\bar{\eta} = 0$. The curves are terminated at the predicted macroscopic fracture initiation strain $\langle \bar{\varepsilon}_\mathrm{f} \rangle$ (indicated using a marker).}
  \label{fig:macroscopic_parameters_phivar}
\end{figure}

The predicted macroscopic stress--strain response in pure shear ($\bar{\eta} = 0$) is shown in Figure~\ref{fig:macroscopic_parameters_phivar} for different hard phase volume fractions $\varphi^\mathrm{hard}$, where the fracture initiation is indicated with a marker. The increase of the hard phase volume fraction results in an increase of the macroscopic yield strength while the macroscopic hardening is more or less constant. The macroscopic fracture initiation strain $\langle \bar{\varepsilon}_\mathrm{f} \rangle$ decreases with increasing hard phase volume fraction (as already observed in Figure~\ref{fig:epsf_parameters_phivar}) while at the same time the fracture initiation stress $\langle \bar{\tau}_\mathrm{eq}^\mathrm{f} \rangle$ increases.

\begin{figure}[htp]
  \centering
  \includegraphics[width=.5\linewidth]{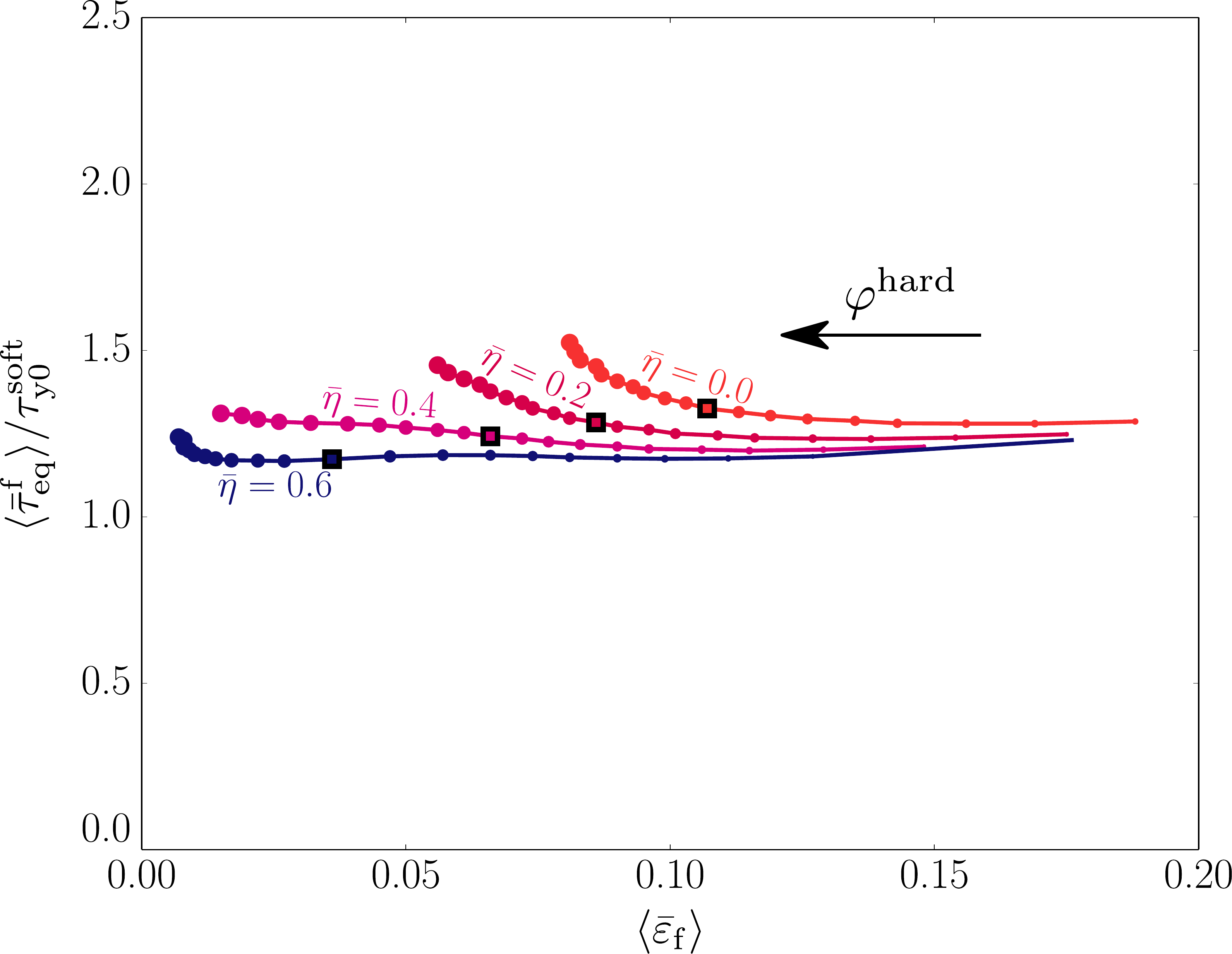}
  \caption{Predicted fracture initiation equivalent stress $\langle \bar{\tau}_\mathrm{eq}^\mathrm{f} \rangle$ and equivalent strain $\langle \bar{\varepsilon}_\mathrm{f} \rangle$ for varying hard phase volume fractions $\varphi^\mathrm{hard}$ (indicated with an arrow) at constant phase contrast $\chi = 2$. The different curves corresponds to different macroscopic stress triaxialities $\bar{\eta}$. The reference ensemble ($\varphi^\mathrm{hard} = 0.25$ and $\chi = 2$) is highlighted using a black square.}
  \label{fig:macroscopic_eta_phivar}
\end{figure}

The consequence of the above observations is examined next in the context of the classical trade-off between strength and ductility. In Figure~\ref{fig:macroscopic_eta_phivar}, the macroscopic fracture initiation stress, $\langle \bar{\tau}_\mathrm{eq}^\mathrm{f} \rangle$, is plotted versus the fracture initiation strain $\langle \bar{\varepsilon}_\mathrm{f} \rangle$ for different macroscopic triaxialities $\bar{\eta}$. The arrow and the increasing size of the markers indicate the increase of the hard phase volume fraction. For $\bar{\eta} = 0$ it is observed, as in Figure~\ref{fig:macroscopic_parameters_phivar}, that for an increasing hard phase volume fraction, the strength, $\langle \bar{\tau}_\mathrm{eq}^\mathrm{f} \rangle$, increases at the expense of decreasing ductility, $\langle \bar{\varepsilon}_\mathrm{f} \rangle$. However, at the highest considered triaxiality of $\bar{\eta} = 0.6$ this trade-off breaks down. At this triaxiality, an increase of the hard phase volume fraction causes the fracture to be dominated by the hard phase, leading to an almost constant strength and decreasing ductility. In line with Figure~\ref{fig:epsf_parameters_phivar}, this is a direct consequence of the competition between failure dominated by either the soft phase or the hard phase, whereby the latter `wins' for $\bar{\eta} = 0.6$.

\section{Influence of phase contrast}
\label{sec:chi}

Next, the phase contrast $\chi$ is varied using different\footnote{The same set of microstructures is used, the only difference is the yielding.} ensembles with a constant hard phase volume fraction of $\varphi^\mathrm{hard} = 0.25$.

\begin{figure}[htp]
  \centering
  \includegraphics[width=.5\textwidth]{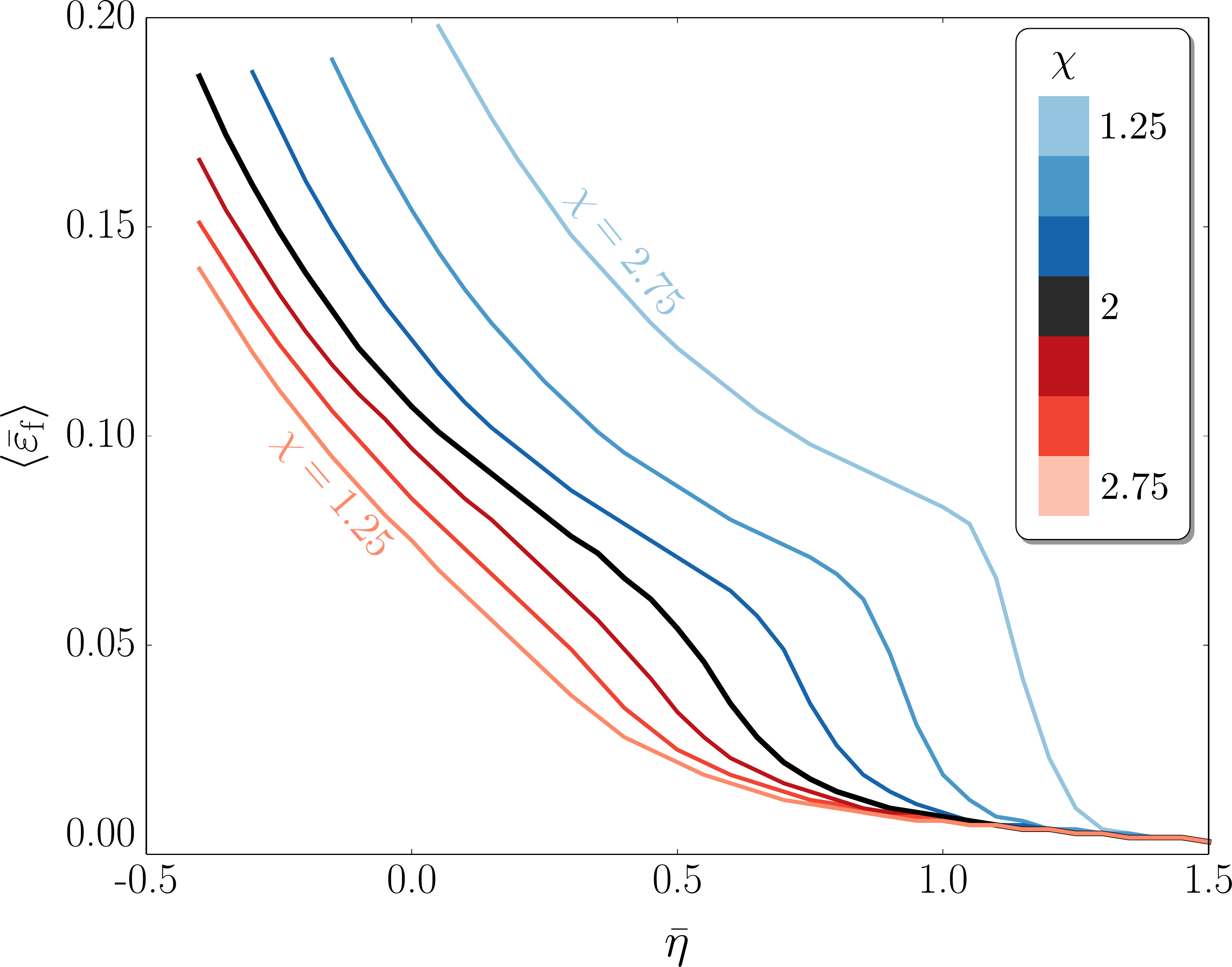}
  \caption{Macroscopic fracture initiation strain $\langle \bar{\varepsilon}_\mathrm{f} \rangle$ as a function of the applied triaxiality $\bar{\eta}$, for different yield stress ratios between the hard and the soft phase, $\chi$, at constant hard phase volume fraction $\varphi^\mathrm{hard} = 0.25$, cf.\ Figure~\ref{fig:epsf_parameters_phivar}.}
  \label{fig:epsf_parameters_chivar}
\end{figure}

The macroscopic fracture initiation strain $\langle \bar{\varepsilon}_\mathrm{f} \rangle$ as a function of macroscopic stress triaxiality $\bar{\eta}$ is shown in Figure~\ref{fig:epsf_parameters_chivar} for seven different values of $\chi$. Both the soft phase and the hard phase failure mechanisms are strongly promoted by increasing the phase contrast, however at a different rate. Consequently, the outcome of the competition between both failure mechanisms is different for different phase contrasts. At low phase contrast (light blue), the fracture initiation strain $\langle \bar{\varepsilon}_\mathrm{f} \rangle$ is significantly higher than for the reference ensemble (in black) as it is dominated by ductile fracture initiation for almost all considered triaxialities. At higher phase contrast (in red) the fracture is entirely in the hard phase fracture regime.

\begin{figure}[htp]
  \centering
  \includegraphics[width=.5\linewidth]{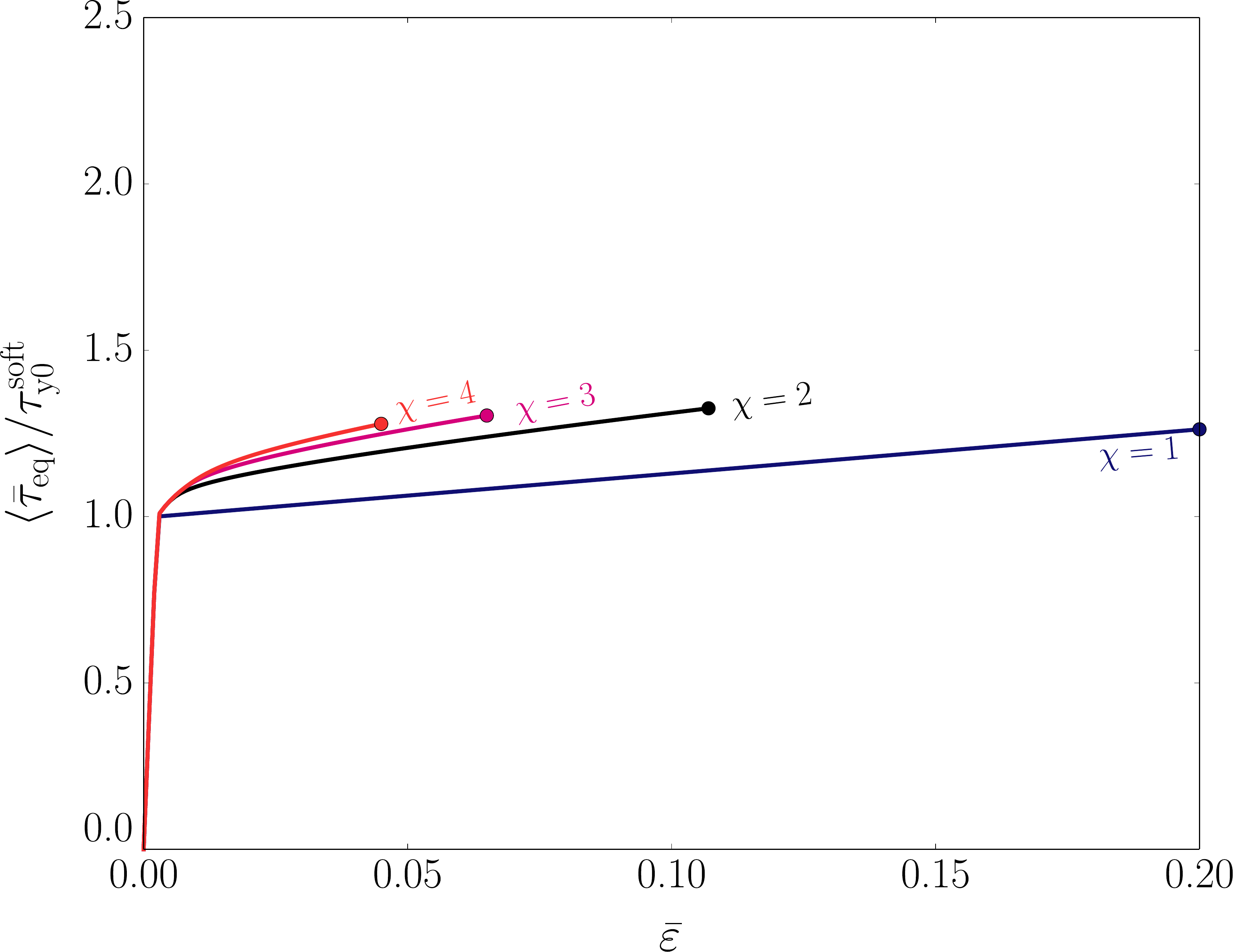}
  \caption{The ensemble averaged macroscopic equivalent stress $\langle \bar{\tau}_\mathrm{eq} \rangle$ as a function the macroscopic, applied, strain $\bar{\varepsilon}$ for different phase contrast $\chi$ at constant volume fraction $\varphi^\mathrm{hard} = 0.25$; cf.\ Figure~\ref{fig:macroscopic_parameters_phivar}.}
  \label{fig:macroscopic_parameters_chivar}
\end{figure}

The predicted macroscopic stress--strain responses in pure shear are shown in Figure~\ref{fig:macroscopic_parameters_chivar}. The observed trend is quite different from that in Figure~\ref{fig:macroscopic_parameters_phivar}: the yield stress is constant for all $\chi$ while the hardening increases. This increase is much lower than the increase in phase contrast $\chi$ as the soft phase accommodates most of the plastic deformation. Again, the predicted macroscopic fracture initiates at lower strain with increasing $\chi$. Note that the resulting fracture stress $\langle \bar{\tau}_\mathrm{eq} \rangle$ is almost constant due to the rapid decay in fracture strain caused by the different outcome of the competition between failure mechanisms.

\begin{figure}[htp]
  \centering
  \includegraphics[width=.5\linewidth]{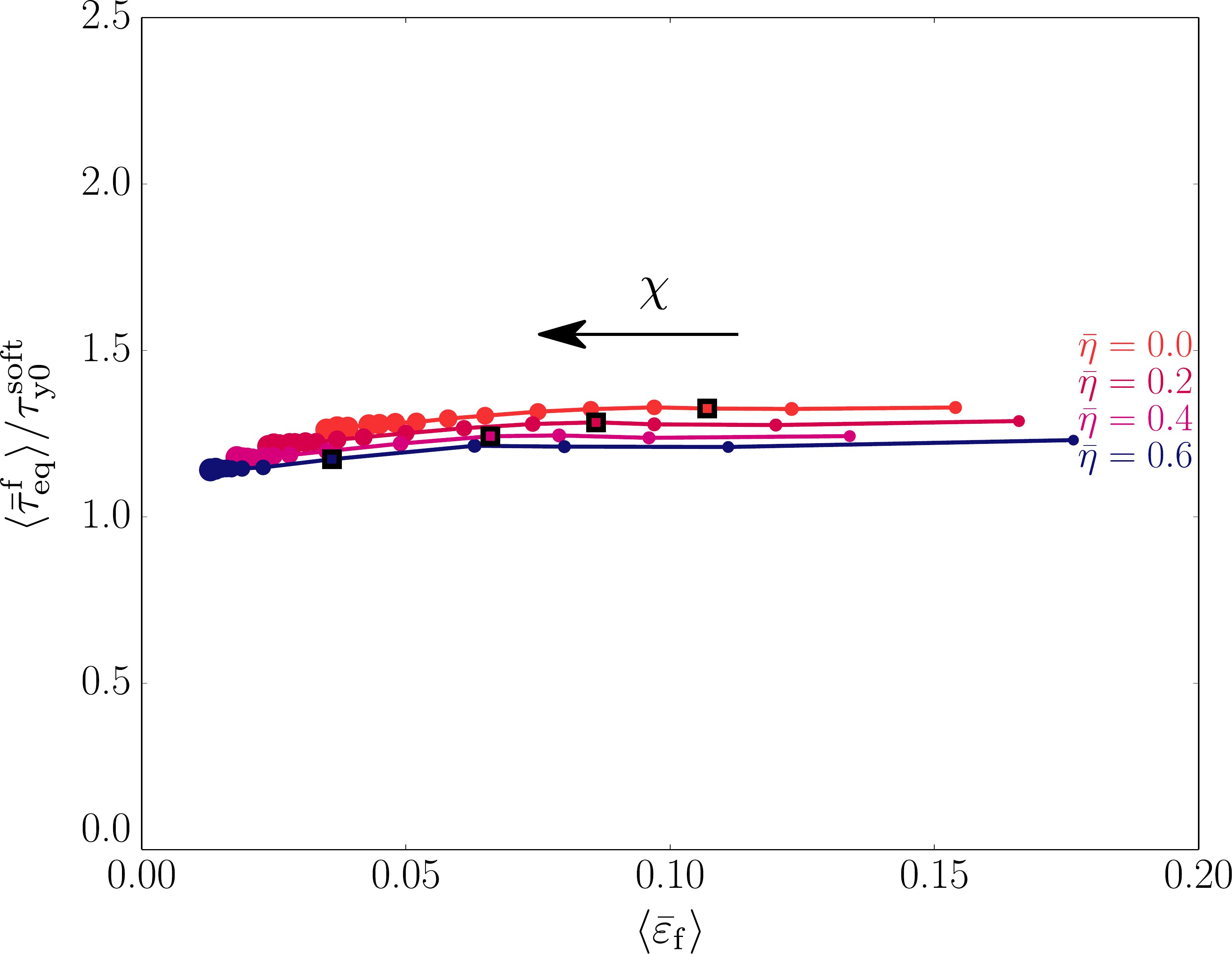}
  \caption{Predicted fracture initiation equivalent stress $\langle \bar{\tau}_\mathrm{eq}^\mathrm{f} \rangle$ and equivalent strain $\langle \bar{\varepsilon}_\mathrm{f} \rangle$ for varying phase contrast $\chi$ (indicated with an arrow) at constant hard phase volume fraction $\varphi^\mathrm{hard} = 0.25$; cf.\ Figure~\ref{fig:macroscopic_eta_phivar}.}
  \label{fig:macroscopic_eta_chivar}
\end{figure}

Figure~\ref{fig:macroscopic_eta_chivar} shows the strength versus ductility trade-off with contrast. For this case, a more or less constant strength is observed accompanied by a decreasing ductility, regardless of the triaxiality. Recall from Figure~\ref{fig:epsf_parameters_chivar} that the increase in phase contrast at this volume fraction leads to hard phase dominated failure. Furthermore, an increase in phase contrast $\chi$ mainly leads to an increase in hardening, not in the yield point (see Figure~\ref{fig:macroscopic_parameters_chivar}). As the fracture strain $\langle \bar{\varepsilon}_\mathrm{f} \rangle$ rapidly decreases and approaches the yield strain, the effect of the increase in hardening is negligible.

\section{Mechanism map}
\label{sec:mechanism}

\subsection{Results}

To understand the consequences of the competition between the two failure mechanisms, the outcomes of all combinations of hard phase volume fractions, phase contrasts, and stress triaxialities are collected in a mechanism map. This map is constructed using approximately $90000$ finite element calculations. Due to the model's simplicity such a computation becomes feasible even on moderately sized computing cluster. Using an in-house code each volume element takes between two and three minutes to compute.

\begin{figure}[htp]
  \centering
  \includegraphics[width=0.5\linewidth]{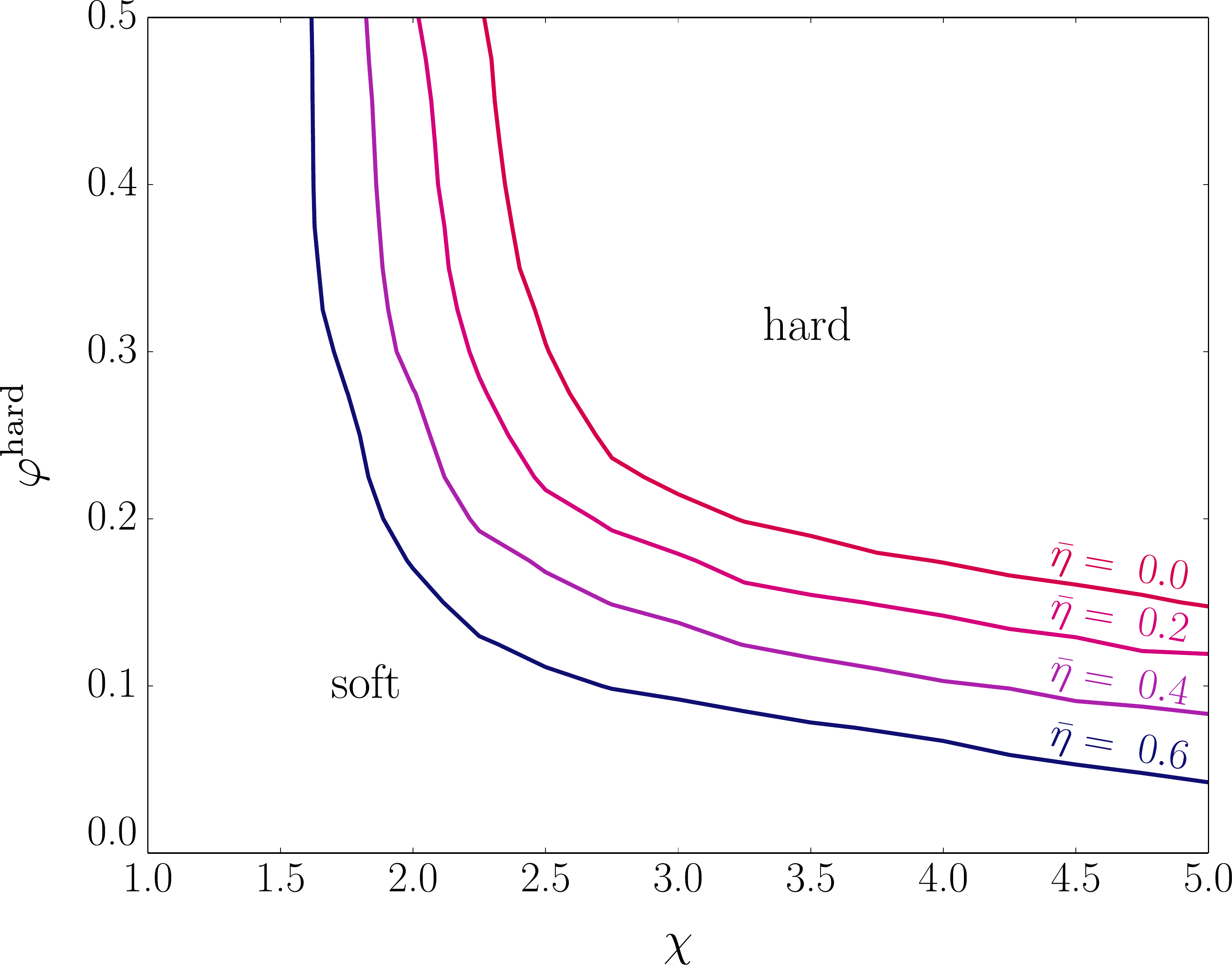}
  \caption{Competition between the initiation of ductile fracture in the soft phase, and brittle fracture in the hard phase, in the form of a mechanism map. At the iso-lines (corresponding to different levels of applied stress triaxiality $\bar{\eta}$) precisely 50 percent of the failed cells are soft and 50 percent are hard. The hard phase volume fraction $\varphi^\mathrm{hard}$ increases along the vertical axis, and the phase contrast $\chi$ along the horizontal axis.}
  \label{fig:mechansim}
\end{figure}

The resulting mechanism map is presented in Figure~\ref{fig:mechansim}. The different curves represent the combination of volume fractions $\varphi^\mathrm{hard}$ and phase contrast $\chi$ for which exactly 50\% of the failed cells are soft and 50\% are hard, at a given value of applied triaxiality $\bar{\eta}$. In the region to the lower left of the curve, that is denoted by ``soft'', the majority of fractured cells are soft, whereas to the upper right (denoted by ``hard'') the majority is hard.

The result is first discussed for a triaxiality $\bar{\eta} = 0$. For a low hard phase volume fraction, $\varphi^\mathrm{hard} < 0.15$, the initiation of fracture is dominated by the soft phase regardless of the contrast in plastic properties between the hard and the soft phase. In this regime the effect of the local incompatibility in deformability remains small as the hard phase is sufficiently dispersed in the microstructure. At the other extreme, where $\varphi^\mathrm{hard} > 0.4$, the initiation of fracture is dominated by the hard phase if the yield contrast is sufficiently large ($\chi > 2.5$). In this regime, the incompatibility between the two phases has a strong influence, as the hard phase domains are close and link up. The transition between these two regimes occurs at approximately $\varphi^\mathrm{hard} = 0.25$. These observations also apply for the other triaxialities although the transition shifts to favor hard phase failure at higher triaxialities as observed before.

\subsection{Discussion}

The outcome of the simple numerical model in the form of this mechanism map is consistent with experimental observations from the literature. For instance, Mummery and Derby \cite{Mummery1991} observe that the  fracture initiation is dominated by breaking of the brittle reinforcement particles for higher volume fractions. And for higher phase contrasts, Lee et al.\ \cite{Lee2004} observe cleavage fracture.

\section{Influence of the local phase distribution}
\label{sec:hotspot}

\subsection{Analysis}

To reveal the correlation between the local microstructural morphology and the individual fracture initiation sites, the average microstructure around the fracture initiation sites is calculated. This approach was first introduced by De Geus et al.\ \cite{DeGeus2015a} and is summarized below. It quantifies the probability of finding the hard phase at a certain position relative to the fracture initiation sites. A value higher than the hard phase volume fraction, $\varphi^\mathrm{hard}$, corresponds to a positive correlation between fracture initiation and hard phase at that relative position. Vice versa a value lower than $\varphi^\mathrm{hard}$ corresponds to a positive correlation between fracture initiation and the soft phase.

The method is discussed based on a single volume element; the ensemble average trivially follows and is therefore omitted. The microstructure is described using a phase indicator $\mathcal{I}$ which is defined as follows:
\begin{equation}
\label{eq:model:I}
  \mathcal{I}(i,j) =
  \begin{cases}
    0 &\quad\mathrm{for}\, (i,j) \in \mathrm{soft} \\
    1 &\quad\mathrm{for}\, (i,j) \in \mathrm{hard}
  \end{cases}
\end{equation}
whereby $(i,j)$ represents the position of a particular cell. For the regular grid used here, it corresponds to the `pixel' position, i.e.\ the row and column index of the grid of square cells.

The average microstructure at a certain position $(\Delta i, \Delta j)$ relative to a fracture initiation site is calculated by averaging the phase indicator $\mathcal{I}$ weighted by the fracture initiation indicator $\mathcal{D}$ over all positions $(i,j)$. I.e.\
\begin{equation} \label{eq:model:hotpot}
  \mathcal{I}_\mathcal{D} (\Delta i, \Delta j) =
  \frac{
    \sum_{ij} \mathcal{D}(i,j) \, \mathcal{I}(i+\Delta i, j+\Delta j)
  }{
    \sum_{ij} \mathcal{D}(i,j) \hfill
  }
\end{equation}
The index $(i,j)$ loops of the cells (or pixels) in the volume element.

Finally, the ensemble average $\langle \mathcal{I}_\mathcal{D} \rangle$ is calculated by averaging over all $256$ volume elements in the ensemble.

\subsection{Results}

\begin{figure}[tph]
  \centering
  \includegraphics[width=.6\linewidth]{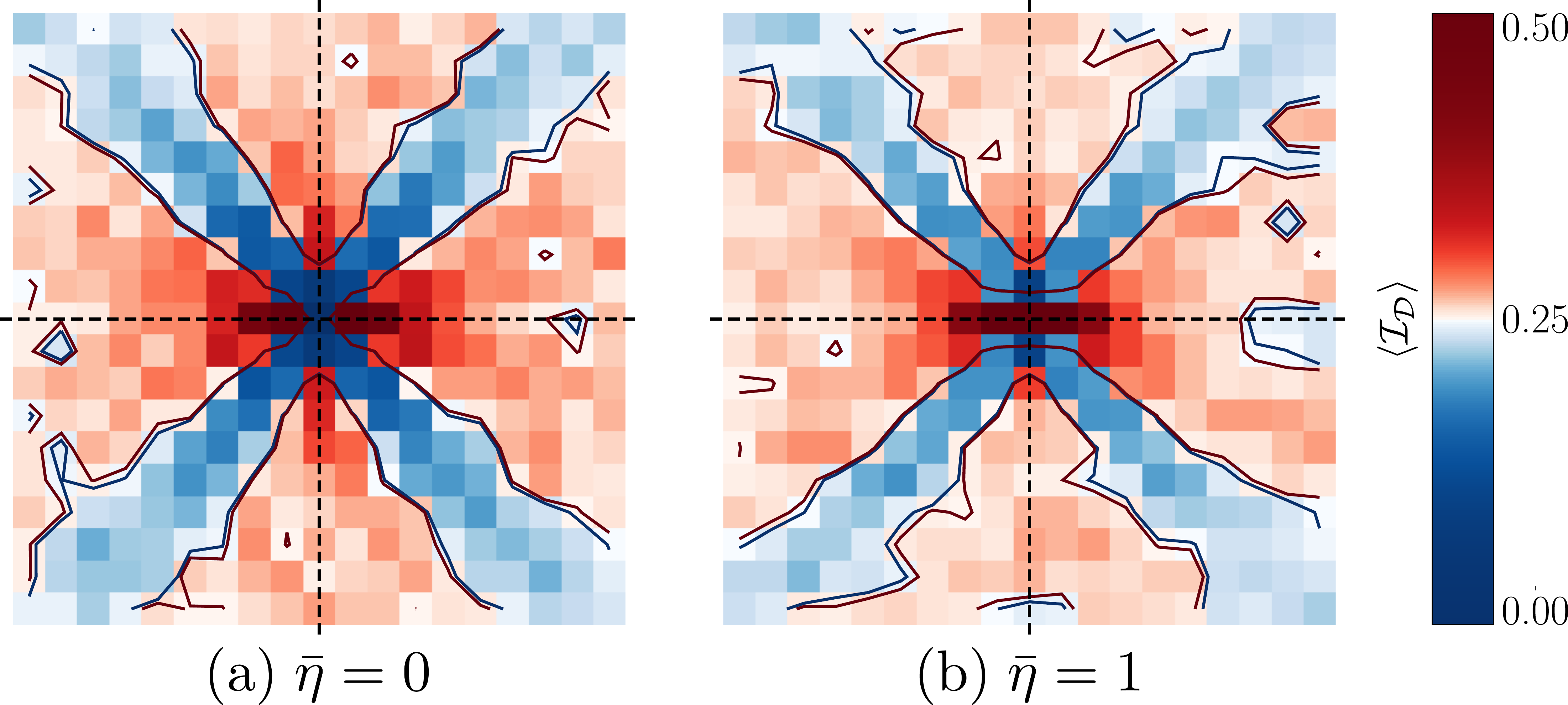}
  \caption{Average microstructure around the fracture initiation sites is shown for different applied stress triaxialities: (a) $\bar{\eta} = 0$ and (b) $\bar{\eta} = 1$; both taken at their respective macroscopic fracture initiation strains $\langle \bar{\varepsilon}_\mathrm{f} \rangle$. }
  \label{fig:hotspot_brittle}
\end{figure}

The result is shown in Figure~\ref{fig:hotspot_brittle} for two different values of the applied triaxiality -- one in the regime where failure is governed by the soft phase ($\bar{\eta} = 0$) and the other in the hard phase failure regime ($\bar{\eta} = 1$). For both, the average arrangement of phases is computed at their respective fracture initiation strain (see Figure~\ref{fig:epsf_reference}). The origin, indicated with black dashed lines, corresponds to the fracture initiation site. The colors are chosen to maximize interpretation: red corresponds to an elevated probability of the hard phase, and blue to that of the soft phase. For $\bar{\eta} = 0$, in Figure~\ref{fig:hotspot_brittle}(a), it observed that, as expected, the fracture initiates in the soft phase: $\langle \mathcal{I}_\mathcal{D} \rangle \approx 0$ at the origin of the diagram. Directly to the left and right, in the tensile direction, regions of the hard phase are found. This band of hard phase is interrupted by bands of the soft phase under angles close to $\pm 45$ degrees. This observation coincides with the result of \cite{DeGeus2015a}, which however was limited to fracture initiation of the soft phase only.

When the result is compared to $\langle \mathcal{I}_D \rangle$ taken at $\bar{\eta} = 1$, in Figure~\ref{fig:hotspot_brittle}(b), it appears that the key features do not change, except that fracture now initiates in the hard phase (observed as $\langle \mathcal{I}_\mathcal{D} > \varphi^\mathrm{hard}$ in the center). Fracture thus initiates in a band of the hard phase aligned with the tensile direction (in red) there where it intersects with a band of the soft phase under an angle of close to $\pm 45$ degrees with respect to the tensile direction (in blue); in this case the hard phase band is not interrupted. This result is remarkable as both results in Figure~\ref{fig:hotspot_brittle} are dominated by totally different failure mechanisms. The only noticeable difference is that the probability of the regions of the soft phase is reduced (characterized by the intensity of blue in the bands at $\pm 45$ degree angles) with respect to Figure~\ref{fig:hotspot_brittle}(a).

\subsection{Discussion}

The above result suggests that a region or band of hard phase aligned with the tensile direction, intersected by regions or bands of the soft phase in the direction of shear are critical for fracture initiation, regardless of whether the hard band is actually interrupted or not. Such characteristics have frequently been observed experimentally \cite{Lewandowski1989,Kim1981,Kim2000,Avramovic-Cingara2009,Kadkhodapour2011,Tasan2010,DeGeus2016,Bareggi2012,Manigandan2012}, although experimentally the actual phase in which fracture initiates often can not be uniquely determined. Based on Figure~\ref{fig:hotspot_brittle} the surroundings would look similar for both fracture initiation mechanisms. Using a numerical model, Segurado and LLorca \cite{Segurado2006, LLorca2004} made the observation that damage nucleation is promoted by clustering of the reinforcement phase in the tensile direction for either hard phase damage, soft phase damage, or interface decohesion. The results in Figure~\ref{fig:hotspot_brittle} are in accordance with these observations, the main difference of the current result is that average microstructure is obtained in a much wider region around the fracture initiation sites. For three-dimensional microstructures, using a similar analysis as presently presented, De Geus et al.\ \citep{DeGeus2016a} have found a qualitatively similar average phase distribution around fracture initiation as Figure~\ref{fig:hotspot_brittle} for a planar deformation (i.e.\ pure shear).

De Geus et al.\ \cite{DeGeus2015a} considered ductile failure in the soft phase only and reasoned that the arrangement of phases in Figure~\ref{fig:hotspot_brittle}(a) is due to (i) a phase boundary perpendicular to the tensile axis giving rise to hydrostatic stress and (ii) shear bands through the soft phase aligned with the shear axis, giving rise to high plastic deformation. From the current result, in Figure~\ref{fig:hotspot_brittle}(b), it is observed that a combination of these mechanisms is also responsible for the high stress in the hard phase. In any case, critical for damage is (i) a band of hard phase aligned with the tensile direction with which (ii) bands of the soft phase that aligned with shear directions intersect. The orientation is determined by the applied macroscopic deformation \citep{DeGeus2016a}. If such a configuration could be avoided, the material's fracture properties would be enhanced.

\section{Comparison with ductile failure of the hard phase}
\label{sec:ductile}

For the hard phase the Rankine model for cleavage has been used so far. It compares the maximum principal stress to a critical value, see equation~\eqref{eq:model:D_hard}, and it is therefore strongly dependent on the actual stress state. Since a small amount of plasticity is allowed in the hard phase and micrographs of the fracture surface often reveal dimples, it may be appropriate to consider also the hard phase ductile. In this section, the Rankine model is therefore replaced by the Johnson-Cook model according to \eqref{eq:model:D_soft}, and both phases thus follow this criterion albeit with different parameter sets. The parameters of the soft phase are like before; those of the hard phase are selected such that the fracture strain is obtained for the homogeneous hard phase in uniaxial tension as for the Rankine-based model used so far. Specifically:
\begin{equation}
  A^\mathrm{hard} = 0.87 \qquad
  B^\mathrm{hard} = 5.6  \qquad
  \varepsilon_\mathrm{pc}^\mathrm{hard} = 0.0
\end{equation}

\begin{figure}[htp]
  \centering
  \includegraphics[width=.5\textwidth]{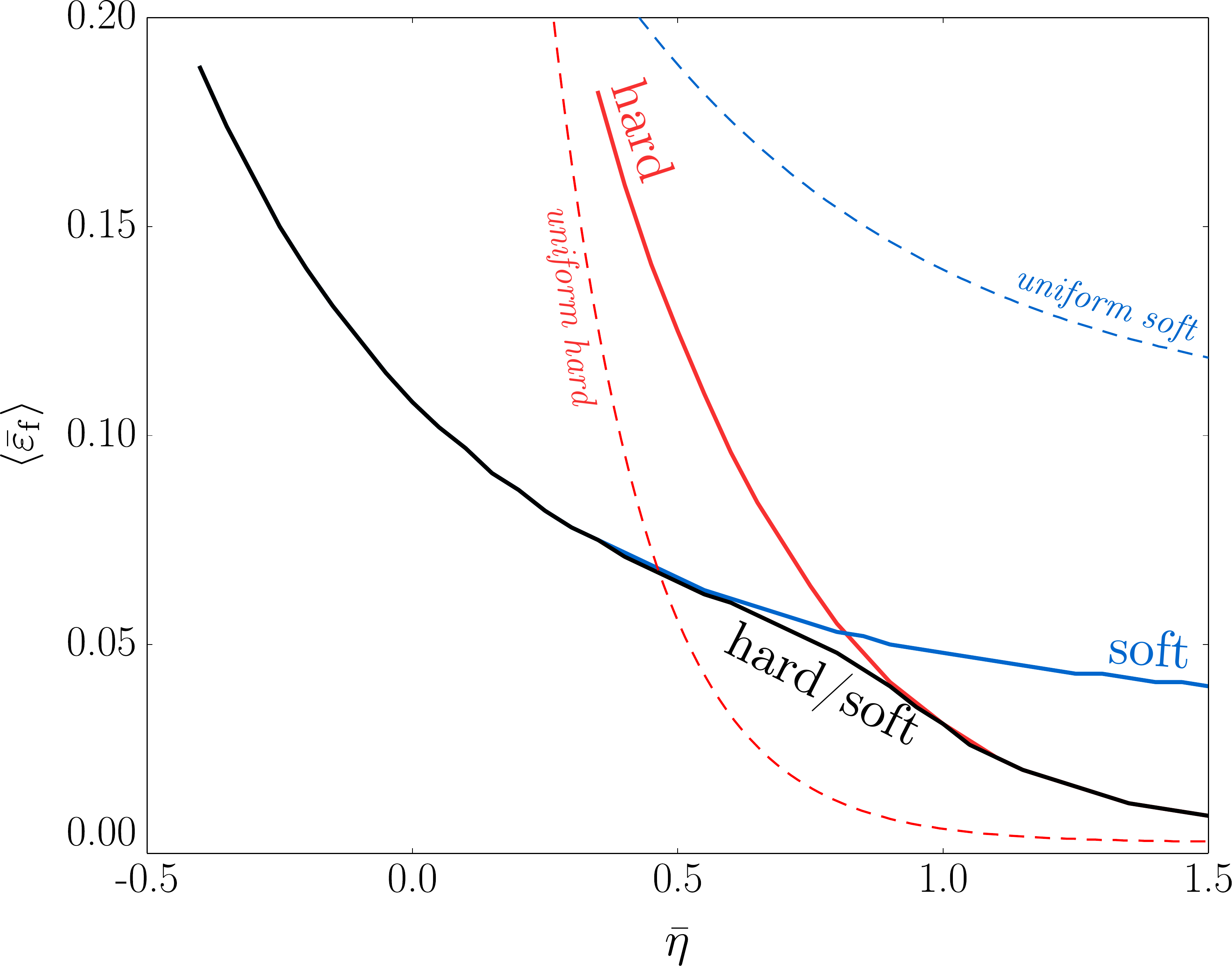}
  \caption{Macroscopic fracture initiation strain $\langle \bar{\varepsilon}_\mathrm{f} \rangle$ as a function of the applied triaxiality $\bar{\eta}$ for the reference configuration, predicted using a ductile damage indicator in both phases; cf.\ Figure~\ref{fig:epsf_reference}.}
  \label{fig:epsf_ductile}
\end{figure}

The macroscopic fracture initiation strain $\langle \bar{\varepsilon}_\mathrm{f} \rangle$ is shown in Figure~\ref{fig:epsf_ductile} as a function of the applied triaxiality $\bar{\eta}$ for the reference parameters in black (i.e.\ hard phase volume fraction $\varphi^\mathrm{hard} = 0.25$ and phase contrast $\chi = 2$). The curves for failure in only one of the phases are included using a red line (only hard phase failure) and blue line (only soft phase failure). The fracture initiation strain for the uniform phases are included as dashed lines. When the black curve is compared to the result in Figure~\ref{fig:epsf_reference}, it appears that the macroscopic fracture initiation strain decreases less strongly with triaxiality. The transition from fracture initiation dominated by the soft phase to that dominated by the hard phase is still observed, however at a different triaxiality. Compared to the uniform phases the same observations can be made as for Figure~\ref{fig:epsf_reference}.

\begin{figure}[tph]
  \centering
  \includegraphics[width=.6\linewidth]{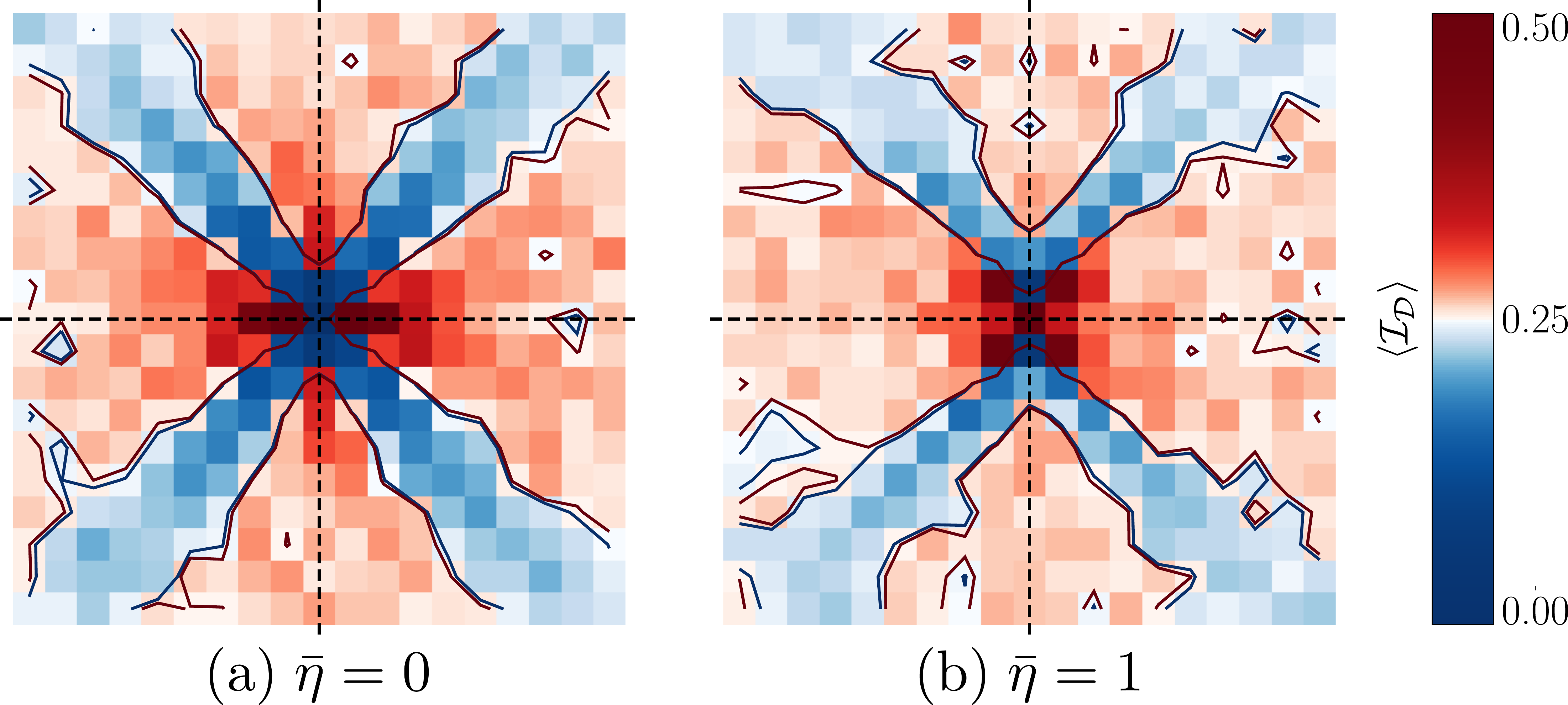}
  \caption{Average microstructure around the fracture initiation sites is shown for an applied stress triaxiality of (a) $\bar{\eta} = 0$ and (b) $\bar{\eta} = 1$.}
  \label{fig:hotspot_ductile}
\end{figure}

The average microstructure around the fracture initiation sites is shown in Figure~\ref{fig:hotspot_ductile}. For $\bar{\eta} = 0$ the result still coincides with Figure~\ref{fig:hotspot_brittle}(a) as fracture initiation is dominated by the soft phase there. Also at $\bar{\eta} = 1$, where the failure is dominated by the hard phase, the key features coincide with Figure~\ref{fig:hotspot_brittle}(b). However, the regions of the hard phase in the direction of tension (to the left and right of the fracture initiation sites) have a lower probability. Oppositely, the regions of the soft phase in the direction of shear ($\pm 45$ degrees) have a higher probability. I.e.\ the relative position of the soft phase around the fracture initiation site is more important, while the relative position of the hard phase is less important compared to the case of cleavage fracture. Also, the orientation of the regions of the soft phase deviates more from $\pm 45$ degrees with respect to the tensile direction. This is due to the different weighing of the plastic strain and the volumetric stress by replacing the Rankine model with the Johnson-Cook model.

\section{Concluding remarks}
\label{sec:conclusion}

A simple multi-scale model is proposed and exploited that uses a microscopic model restricted to the most essential micromechanics underlying the initiation of failure in a ductile two-phase material comprising a hard but brittle phase embedded in a soft and ductile matrix. A large number of microstructures are considered to capture the -- statistically extreme -- fracture initiation at the level of the individual grains. Using this model, the common observations in the literature on the effects of applied stress triaxiality, hard phase volume fraction, and contrast in mechanical properties are analyzed and refined.

\begin{itemize}
  \item The large number of realizations enables the characterization of the average microstructural morphology around the initiation of fracture. Around fracture initiation sites, a band of hard phase is identified in the tensile direction, intersected by bands of soft phase in the directions of shear. This configuration is critical, regardless of whether the band of hard phase is interrupted or not; i.e.\ regardless if the local fracture initiates in the soft phase or in the hard phase.
  \item The macroscopic fracture initiation is dominated by the soft phase at low values of applied stress triaxiality. Above a certain critical triaxiality the balance tips to the hard phase. This critical triaxiality is a function of the hard phase volume fraction and the contrast in mechanical properties between the phases, as characterized by the mechanism map of Figure~\ref{fig:mechansim}.
  \item The increase in hard phase volume fraction leads to an increase in strength at the expense of ductility. Due to the different effect on the macroscopic elasto-plastic behavior, an increase in phase contrast does not lead to a significant increase in strength, while the ductility decreases.
\end{itemize}

It was shown that these conclusions are relatively insensitive the exact value of the parameters. They are thus representative for different materials in this class.

\section*{Acknowledgments}

This research was carried out under project number M22.2.11424 in the framework of the research program of the Materials innovation institute M2i (\href{http://www.m2i.nl}{www.m2i.nl}).

\bibliography{library}

\begin{thebibliography}{65}
\providecommand{\natexlab}[1]{#1}
\providecommand{\url}[1]{\texttt{#1}}
\expandafter\ifx\csname urlstyle\endcsname\relax
  \providecommand{\eprint}[1]{eprint: #1}\else
  \providecommand{\eprint}{eprint: \begingroup \urlstyle{rm}\Url}\fi
\expandafter\ifx\csname urlstyle\endcsname\relax
  \providecommand{\doi}[1]{doi: #1}\else
  \providecommand{\doi}{doi: \begingroup \urlstyle{rm}\Url}\fi

\bibitem[Argon et~al.(1975)Argon, Im, and Safoglu]{Argon1975}
A.S.\ Argon, J.\ Im, and R.\ Safoglu.
\newblock {Cavity formation from inclusions in ductile fracture}.
\newblock \emph{Metall.\ Trans.\ A}, 6\penalty0 (4):\penalty0 825--837, 1975.
\newblock \doi{10.1007/BF02672306}.

\bibitem[Beremin(1981)]{Beremin1981}
F.M.\ Beremin.
\newblock {Cavity formation from inclusions in ductile fracture of A508 steel}.
\newblock \emph{Metall.\ Trans.\ A}, 12\penalty0 (5):\penalty0 723--731, 1981.
\newblock \doi{10.1007/BF02648336}.

\bibitem[Bao et~al.(1991)Bao, Hutchinson, and McMeeking]{Bao1991}
G.\ Bao, J.W.\ Hutchinson, and R.M.\ McMeeking.
\newblock {Particle reinforcement of ductile matrices against plastic flow and
  creep}.
\newblock \emph{Acta Metall.\ Mater.}, 39\penalty0 (8):\penalty0 1871--1882,
  1991.
\newblock \doi{10.1016/0956-7151(91)90156-U}.

\bibitem[Johnson and Cook(1985)]{Johnson1985}
G.R.\ Johnson and W.H.\ Cook.
\newblock {Fracture characteristics of three metals subjected to various
  strains, strain rates, temperatures and pressures}.
\newblock \emph{Eng.\ Fract.\ Mech.}, 21\penalty0 (1):\penalty0 31--48, 1985.
\newblock \doi{10.1016/0013-7944(85)90052-9}.

\bibitem[Choi et~al.(2009)Choi, Liu, Sun, and Khaleel]{Choi2009}
K.S.\ Choi, W.N.\ Liu, X.\ Sun, and M.A.\ Khaleel.
\newblock {Influence of Martensite Mechanical Properties on Failure Mode and
  Ductility of Dual-Phase Steels}.
\newblock \emph{Metall.\ Mater.\ Trans.\ A}, 40\penalty0 (4):\penalty0
  796--809, 2009.
\newblock \doi{10.1007/s11661-009-9792-6}.

\bibitem[Sun et~al.(2009{\natexlab{a}})Sun, Choi, Liu, and Khaleel]{Sun2009}
X.\ Sun, K.S.\ Choi, W.N.\ Liu, and M.A.\ Khaleel.
\newblock {Predicting failure modes and ductility of dual phase steels using
  plastic strain localization}.
\newblock \emph{Int.\ J.\ Plast.}, 25\penalty0 (10):\penalty0 1888--1909,
  2009{\natexlab{a}}.
\newblock \doi{10.1016/j.ijplas.2008.12.012}.

\bibitem[Deng and Chawla(2006)]{Deng2006}
X.\ Deng and N.\ Chawla.
\newblock {Modeling the effect of particle clustering on the mechanical
  behavior of SiC particle reinforced Al matrix composites}.
\newblock \emph{J.\ Mater.\ Sci.}, 41\penalty0 (17):\penalty0 5731--5734, 2006.
\newblock \doi{10.1007/s10853-006-0100-1}.

\bibitem[Heinrich et~al.(2012)Heinrich, Aldridge, Wineman, Kieffer, Waas, and
  Shahwan]{Heinrich2012}
C.\ Heinrich, M.\ Aldridge, A.S.\ Wineman, J.\ Kieffer, A.M.\ Waas, and K.\
  Shahwan.
\newblock {The influence of the representative volume element (RVE) size on the
  homogenized response of cured fiber composites}.
\newblock \emph{Model.\ Simul.\ Mater.\ Sci.\ Eng.}, 20\penalty0 (7):\penalty0
  075007, 2012.
\newblock \doi{10.1088/0965-0393/20/7/075007}.

\bibitem[Povirk(1995)]{Povirk1995}
G.L.\ Povirk.
\newblock {Incorporation of microstructural information into models of
  two-phase materials}.
\newblock \emph{Acta Metall.\ Mater.}, 43\penalty0 (8):\penalty0 3199--3206,
  1995.
\newblock \doi{10.1016/0956-7151(94)00487-3}.

\bibitem[Scheunemann et~al.(2013)Scheunemann, Balzani, Brands, Schr{\"{o}}der,
  and Raabe]{Scheunemann2013}
L.\ Scheunemann, D.\ Balzani, D.\ Brands, J.\ Schr{\"{o}}der, and D.\ Raabe.
\newblock {Statistically similar RVE construction based on 3D dual-phase steel
  microstructures}.
\newblock In \emph{Res. Appl. Struct. Eng. Mech. Comput.} CRC Press, 2013.
\newblock ISBN 9781138000612.
\newblock \doi{10.1201/b15963-77}.

\bibitem[Mortensen and Llorca(2010)]{Mortensen2010}
A.\ Mortensen and J.\ Llorca.
\newblock {Metal Matrix Composites}.
\newblock \emph{Annu.\ Rev.\ Mater.\ Res.}, 40\penalty0 (1):\penalty0 243--270,
  2010.
\newblock \doi{10.1146/annurev-matsci-070909-104511}.

\bibitem[Rashid(1981)]{Rashid1981}
M.S.\ Rashid.
\newblock {Dual Phase Steels}.
\newblock \emph{Annu.\ Rev.\ Mater.\ Sci.}, 11\penalty0 (1):\penalty0 245--266,
  1981.
\newblock \doi{10.1146/annurev.ms.11.080181.001333}.

\bibitem[Tasan et~al.(2014)Tasan, Diehl, Yan, Bechtold, Roters, Schemmann,
  Zheng, Peranio, Ponge, Koyama, Tsuzaki, and Raabe]{Tasan2014b}
C.C.\ Tasan, M.\ Diehl, D.\ Yan, M.\ Bechtold, F.\ Roters, L.\ Schemmann, C.\
  Zheng, N.\ Peranio, D.\ Ponge, M.\ Koyama, K.\ Tsuzaki, and D.\ Raabe.
\newblock {An Overview of Dual-Phase Steels: Advances in
  Microstructure-Oriented Processing and Micromechanically Guided Design}.
\newblock \emph{Annu.\ Rev.\ Mater.\ Res.}, 45\penalty0 (1):\penalty0
  150504161757009, 2014.
\newblock \doi{10.1146/annurev-matsci-070214-021103}.

\bibitem[Davies(1978)]{Davies1978a}
R.G.\ Davies.
\newblock {Influence of martensite composition and content on the properties of
  dual phase steels}.
\newblock \emph{Metall.\ Trans.\ A}, 9\penalty0 (5):\penalty0 671--679, 1978.
\newblock \doi{10.1007/BF02659924}.

\bibitem[Lawson et~al.(1981)Lawson, Metlock, and Krauss]{Lawson1981}
R.D.\ Lawson, D.K.\ Metlock, and G.\ Krauss.
\newblock {Fundamentals of Dual-Phase Steels}.
\newblock pages 347--381. AIME, New York, NY, 1981.

\bibitem[Speich and Miller(1979)]{Speich1979}
G.R.\ Speich and R.L.\ Miller.
\newblock {Structure and Properties of Dual-Phase Steels}.
\newblock pages 145--182. AIME, New York, NY, 1979.

\bibitem[Ahmad et~al.(2000)Ahmad, Manzoor, Ali, and Akhter]{Ahmad2000}
E.\ Ahmad, T.\ Manzoor, K.L.\ Ali, and J.I.\ Akhter.
\newblock {Effect of microvoid formation on the tensile properties of
  dual-phase steel}.
\newblock \emph{J.\ Mater.\ Eng.\ Perform.}, 9\penalty0 (3):\penalty0 306--310,
  2000.
\newblock \doi{10.1361/105994900770345962}.

\bibitem[Llorca et~al.(1991)Llorca, Needleman, and Suresh]{Llorca1991}
J.\ Llorca, A.\ Needleman, and S.\ Suresh.
\newblock {An analysis of the effects of matrix void growth on deformation and
  ductility in metal-ceramic composites}.
\newblock \emph{Acta Metall.\ Mater.}, 39\penalty0 (10):\penalty0 2317--2335,
  1991.
\newblock \doi{10.1016/0956-7151(91)90014-R}.

\bibitem[{Le Roy} et~al.(1981){Le Roy}, Embury, Edwards, and Ashby]{LeRoy1981}
G.\ {Le Roy}, J.D.\ Embury, G.\ Edwards, and M.F.\ Ashby.
\newblock {A model of ductile fracture based on the nucleation and growth of
  voids}.
\newblock \emph{Acta Metall.}, 29\penalty0 (8):\penalty0 1509--1522, 1981.
\newblock \doi{10.1016/0001-6160(81)90185-1}.

\bibitem[Mummery and Derby(1991)]{Mummery1991}
P.\ Mummery and B.\ Derby.
\newblock {The influence of microstructure on the fracture behaviour of
  particulate metal matrix composites}.
\newblock \emph{Mater.\ Sci.\ Eng.\ A}, 135:\penalty0 221--224, 1991.
\newblock \doi{10.1016/0921-5093(91)90566-6}.

\bibitem[Lee et~al.(2004)Lee, Hwang, Lee, Lee, and Kim]{Lee2004}
H.S.\ Lee, B.\ Hwang, S.\ Lee, C.G.\ Lee, and S.-J.\ Kim.
\newblock {Effects of martensite morphology and tempering on dynamic
  deformation behavior of dual-phase steels}.
\newblock \emph{Metall.\ Mater.\ Trans.\ A}, 35\penalty0 (8):\penalty0
  2371--2382, 2004.
\newblock \doi{10.1007/s11661-006-0217-5}.

\bibitem[Kang et~al.(2007)Kang, Ososkov, Embury, and Wilkinson]{Kang2007}
J.\ Kang, Y.\ Ososkov, J.D.\ Embury, and D.S.\ Wilkinson.
\newblock {Digital image correlation studies for microscopic strain
  distribution and damage in dual phase steels}.
\newblock \emph{Scr.\ Mater.}, 56\penalty0 (11):\penalty0 999--1002, 2007.
\newblock \doi{10.1016/j.scriptamat.2007.01.031}.

\bibitem[Maire et~al.(2007)Maire, Carmona, Courbon, and Ludwig]{Maire2007}
E.\ Maire, V.\ Carmona, J.\ Courbon, and W.\ Ludwig.
\newblock {Fast X-ray tomography and acoustic emission study of damage in
  metals during continuous tensile tests}.
\newblock \emph{Acta Mater.}, 55\penalty0 (20):\penalty0 6806--6815, 2007.
\newblock \doi{10.1016/j.actamat.2007.08.043}.

\bibitem[Cox and Low(1974)]{Cox1974}
T.B.\ Cox and J.R.J.\ Low.
\newblock {An investigation of the plastic fracture of AISI 4340 and 18
  Nickel-200 grade maraging steels}.
\newblock \emph{Metall.\ Trans.\ B}, 5\penalty0 (6):\penalty0 1457--1470, 1974.
\newblock \doi{10.1007/BF02646633}.

\bibitem[Papaefthymiou et~al.(2006)Papaefthymiou, Prahl, Bleck, van~der Zwaag,
  and Sietsma]{Papaefthymiou2006}
S.\ Papaefthymiou, U.\ Prahl, W.\ Bleck, S.\ van~der Zwaag, and J.\ Sietsma.
\newblock {Experimental observations on the correlation between microstructure
  and fracture of multiphase steels}.
\newblock \emph{Int.\ J.\ Mater.\ Res.}, 97\penalty0 (12):\penalty0 1723--1731,
  2006.
\newblock \doi{10.3139/146.101406}.

\bibitem[Uthaisangsuk et~al.(2008)Uthaisangsuk, Prahl, and
  Bleck]{Uthaisangsuk2008}
V.\ Uthaisangsuk, U.\ Prahl, and W.\ Bleck.
\newblock {Micromechanical modelling of damage behaviour of multiphase steels}.
\newblock \emph{Comput.\ Mater.\ Sci.}, 43\penalty0 (1):\penalty0 27--35, 2008.
\newblock \doi{10.1016/j.commatsci.2007.07.035}.

\bibitem[Prahl et~al.(2007)Prahl, Papaefthymiou, Uthaisangsuk, Bleck, Sietsma,
  and van~der Zwaag]{Prahl2007}
U.\ Prahl, S.\ Papaefthymiou, V.\ Uthaisangsuk, W.\ Bleck, J.\ Sietsma, and S.\
  van~der Zwaag.
\newblock {Micromechanics-based modelling of properties and failure of
  multiphase steels}.
\newblock \emph{Comput.\ Mater.\ Sci.}, 39\penalty0 (1):\penalty0 17--22, 2007.
\newblock \doi{10.1016/j.commatsci.2006.01.023}.

\bibitem[Anderson et~al.(2014)Anderson, Winkler, Bardelcik, and
  Worswick]{Anderson2014}
D.\ Anderson, S.\ Winkler, A.\ Bardelcik, and M.J.\ Worswick.
\newblock {Influence of stress triaxiality and strain rate on the failure
  behavior of a dual-phase DP780 steel}.
\newblock \emph{Mater.\ Des.}, 60:\penalty0 198--207, 2014.
\newblock \doi{10.1016/j.matdes.2014.03.073}.

\bibitem[Lewandowski et~al.(1989)Lewandowski, Liu, and Hunt]{Lewandowski1989}
J.J.\ Lewandowski, C.\ Liu, and W.H.J.\ Hunt.
\newblock {Effects of matrix microstructure and particle distribution on
  fracture of an aluminum metal matrix composite}.
\newblock \emph{Mater.\ Sci.\ Eng.\ A}, 107:\penalty0 241--255, 1989.
\newblock \doi{10.1016/0921-5093(89)90392-4}.

\bibitem[Kim and Thomas(1981)]{Kim1981}
N.J.\ Kim and G.\ Thomas.
\newblock {Effects of morphology on the mechanical behavior of a dual phase
  Fe/2Si/0.1C steel}.
\newblock \emph{Metall.\ Trans.\ A}, 12\penalty0 (3):\penalty0 483--489, 1981.
\newblock \doi{10.1007/BF02648546}.

\bibitem[Kim and Lee(2000)]{Kim2000}
S.\ Kim and S.\ Lee.
\newblock {Effects of martensite morphology and volume fraction on quasi-static
  and dynamic deformation behavior of dual-phase steels}.
\newblock \emph{Metall.\ Mater.\ Trans.\ A}, 31\penalty0 (7):\penalty0
  1753--1760, 2000.
\newblock \doi{10.1007/s11661-998-0328-2}.

\bibitem[Avramovic-Cingara et~al.(2009{\natexlab{a}})Avramovic-Cingara,
  Ososkov, Jain, and Wilkinson]{Avramovic-Cingara2009}
G.\ Avramovic-Cingara, Y.\ Ososkov, M.K.\ Jain, and D.S.\ Wilkinson.
\newblock {Effect of martensite distribution on damage behaviour in DP600 dual
  phase steels}.
\newblock \emph{Mater.\ Sci.\ Eng.\ A}, 516\penalty0 (1-2):\penalty0 7--16,
  2009{\natexlab{a}}.
\newblock \doi{10.1016/j.msea.2009.03.055}.

\bibitem[Kadkhodapour et~al.(2011)Kadkhodapour, Butz, and
  Ziaei-Rad]{Kadkhodapour2011}
J.\ Kadkhodapour, A.\ Butz, and S.\ Ziaei-Rad.
\newblock {Mechanisms of void formation during tensile testing in a commercial,
  dual-phase steel}.
\newblock \emph{Acta Mater.}, 59\penalty0 (7):\penalty0 2575--2588, 2011.
\newblock \doi{10.1016/j.actamat.2010.12.039}.

\bibitem[Tasan et~al.(2010)Tasan, Hoefnagels, and Geers]{Tasan2010}
C.C.\ Tasan, J.P.M.\ Hoefnagels, and M.G.D.\ Geers.
\newblock {Microstructural banding effects clarified through micrographic
  digital image correlation}.
\newblock \emph{Scr.\ Mater.}, 62\penalty0 (11):\penalty0 835--838, 2010.
\newblock \doi{10.1016/j.scriptamat.2010.02.014}.

\bibitem[Kumar et~al.(2006)Kumar, Briant, and Curtin]{Kumar2006}
H.\ Kumar, C.L.\ Briant, and W.A.\ Curtin.
\newblock {Using microstructure reconstruction to model mechanical behavior in
  complex microstructures}.
\newblock \emph{Mech.\ Mater.}, 38\penalty0 (8-10):\penalty0 818--832, 2006.
\newblock \doi{10.1016/j.mechmat.2005.06.030}.

\bibitem[Vajragupta et~al.(2012)Vajragupta, Uthaisangsuk, Schmaling,
  M{\"{u}}nstermann, Hartmaier, and Bleck]{Vajragupta2012}
N.\ Vajragupta, V.\ Uthaisangsuk, B.\ Schmaling, S.\ M{\"{u}}nstermann, A.\
  Hartmaier, and W.\ Bleck.
\newblock {A micromechanical damage simulation of dual phase steels using
  XFEM}.
\newblock \emph{Comput.\ Mater.\ Sci.}, 54:\penalty0 271--279, 2012.
\newblock \doi{10.1016/j.commatsci.2011.10.035}.

\bibitem[de~Geus et~al.(2015{\natexlab{a}})de~Geus, Peerlings, and
  Geers]{DeGeus2015a}
T.W.J.\ de~Geus, R.H.J.\ Peerlings, and M.G.D.\ Geers.
\newblock {Microstructural topology effects on the onset of ductile failure in
  multi-phase materials -- A systematic computational approach}.
\newblock \emph{Int.\ J.\ Solids Struct.}, 67-68:\penalty0 326--339,
  2015{\natexlab{a}}.
\newblock \doi{10.1016/j.ijsolstr.2015.04.035}.
\newblock \eprint{1604.02858}.

\bibitem[Avramovic-Cingara et~al.(2009{\natexlab{b}})Avramovic-Cingara, Saleh,
  Jain, and Wilkinson]{Avramovic-Cingara2009a}
G.\ Avramovic-Cingara, C.A.R.\ Saleh, M.K.\ Jain, and D.S.\ Wilkinson.
\newblock {Void Nucleation and Growth in Dual-Phase Steel 600 during Uniaxial
  Tensile Testing}.
\newblock \emph{Metall.\ Mater.\ Trans.\ A}, 40\penalty0 (13):\penalty0
  3117--3127, 2009{\natexlab{b}}.
\newblock \doi{10.1007/s11661-009-0030-z}.

\bibitem[Segurado et~al.(2003)Segurado, Gonz{\'{a}}lez, and
  Llorca]{Segurado2003}
J.\ Segurado, C.\ Gonz{\'{a}}lez, and J.\ Llorca.
\newblock {A numerical investigation of the effect of particle clustering on
  the mechanical properties of composites}.
\newblock \emph{Acta Mater.}, 51\penalty0 (8):\penalty0 2355--2369, 2003.
\newblock \doi{10.1016/S1359-6454(03)00043-0}.

\bibitem[Williams et~al.(2010)Williams, Flom, Amell, Chawla, Xiao, and {De
  Carlo}]{Williams2010}
J.J.\ Williams, Z.\ Flom, A.A.\ Amell, N.\ Chawla, X.\ Xiao, and F.\ {De
  Carlo}.
\newblock {Damage evolution in SiC particle reinforced Al alloy matrix
  composites by X-ray synchrotron tomography}.
\newblock \emph{Acta Mater.}, 58\penalty0 (18):\penalty0 6194--6205, 2010.
\newblock \doi{10.1016/j.actamat.2010.07.039}.

\bibitem[Williams et~al.(2012)Williams, Segurado, Llorca, and
  Chawla]{Williams2012}
J.J.\ Williams, J.\ Segurado, J.\ Llorca, and N.\ Chawla.
\newblock {Three dimensional (3D) microstructure-based modeling of interfacial
  decohesion in particle reinforced metal matrix composites}.
\newblock \emph{Mater.\ Sci.\ Eng.\ A}, 557:\penalty0 113--118, 2012.
\newblock \doi{10.1016/j.msea.2012.05.108}.

\bibitem[Segurado and Llorca(2006)]{Segurado2006}
J.\ Segurado and J.\ Llorca.
\newblock {Computational micromechanics of composites: The effect of particle
  spatial distribution}.
\newblock \emph{Mech.\ Mater.}, 38\penalty0 (8-10):\penalty0 873--883, 2006.
\newblock \doi{10.1016/j.mechmat.2005.06.026}.

\bibitem[Simo(1992)]{Simo1992a}
J.C.\ Simo.
\newblock {Algorithms for static and dynamic multiplicative plasticity that
  preserve the classical return mapping schemes of the infinitesimal theory}.
\newblock \emph{Comput.\ Methods Appl.\ Mech.\ Eng.}, 99\penalty0 (1):\penalty0
  61--112, 1992.
\newblock \doi{10.1016/0045-7825(92)90123-2}.

\bibitem[Sun et~al.(2009{\natexlab{b}})Sun, Choi, Soulami, Liu, and
  Khaleel]{Sun2009a}
X.\ Sun, K.S.\ Choi, A.\ Soulami, W.N.\ Liu, and M.A.\ Khaleel.
\newblock {On key factors influencing ductile fractures of dual phase (DP)
  steels}.
\newblock \emph{Mater.\ Sci.\ Eng.\ A}, 526\penalty0 (1-2):\penalty0 140--149,
  2009{\natexlab{b}}.
\newblock \doi{10.1016/j.msea.2009.08.010}.

\bibitem[Al-Abbasi and Nemes(2003)]{Al-Abbasi2003}
F.M.\ Al-Abbasi and J.A.\ Nemes.
\newblock {Micromechanical modeling of dual phase steels}.
\newblock \emph{Int.\ J.\ Mech.\ Sci.}, 45\penalty0 (9):\penalty0 1449--1465,
  2003.
\newblock \doi{10.1016/j.ijmecsci.2003.10.007}.

\bibitem[Asgari et~al.(2009)Asgari, Hodgson, Yang, and Rolfe]{Asgari2009}
S.A.\ Asgari, P.D.\ Hodgson, C.\ Yang, and B.F.\ Rolfe.
\newblock {Modeling of Advanced High Strength Steels with the realistic
  microstructure-strength relationships}.
\newblock \emph{Comput.\ Mater.\ Sci.}, 45\penalty0 (4):\penalty0 860--866,
  2009.
\newblock \doi{10.1016/j.commatsci.2008.12.003}.

\bibitem[de~Geus et~al.(2015{\natexlab{b}})de~Geus, Peerlings, and
  Geers]{DeGeus2015}
T.W.J.\ de~Geus, R.H.J.\ Peerlings, and M.G.D.\ Geers.
\newblock {Microstructural modeling of ductile fracture initiation in
  multi-phase materials}.
\newblock \emph{Eng.\ Fract.\ Mech.}, 147:\penalty0 318--330,
  2015{\natexlab{b}}.
\newblock \doi{10.1016/j.engfracmech.2015.04.010}.
\newblock \eprint{1604.03811}.

\bibitem[Bareggi et~al.(2012)Bareggi, Maire, Bouaziz, and {Di
  Michiel}]{Bareggi2012}
A.\ Bareggi, E.\ Maire, O.\ Bouaziz, and M.\ {Di Michiel}.
\newblock {Damage in dual phase steels and its constituents studied by X-ray
  tomography}.
\newblock \emph{Int.\ J.\ Fract.}, 174\penalty0 (2):\penalty0 217--227, 2012.
\newblock \doi{10.1007/s10704-012-9692-4}.

\bibitem[Hoefnagels et~al.(2015)Hoefnagels, Tasan, Maresca, Peters, and
  Kouznetsova]{Hoefnagels2015}
J.P.M.\ Hoefnagels, C.C.\ Tasan, F.\ Maresca, F.J.\ Peters, and V.G.\
  Kouznetsova.
\newblock {Retardation of plastic instability via damage-enabled microstrain
  delocalization}.
\newblock \emph{J.\ Mater.\ Sci.}, 50\penalty0 (21):\penalty0 6882--6897, 2015.
\newblock \doi{10.1007/s10853-015-9164-0}.

\bibitem[Paul(2013)]{Paul2013a}
S.K.\ Paul.
\newblock {Effect of martensite volume fraction on stress triaxiality and
  deformation behavior of dual phase steel}.
\newblock \emph{Mater.\ Des.}, 50:\penalty0 782--789, 2013.
\newblock \doi{10.1016/j.matdes.2013.03.096}.

\bibitem[Bao and Wierzbicki(2004)]{Bao2004a}
Y.\ Bao and T.\ Wierzbicki.
\newblock {A Comparative Study on Various Ductile Crack Formation Criteria}.
\newblock \emph{J.\ Eng.\ Mater.\ Technol.}, 126\penalty0 (3):\penalty0 314,
  2004.
\newblock \doi{10.1115/1.1755244}.

\bibitem[Hauert et~al.(2010)Hauert, Rossoll, and Mortensen]{Hauert2010}
A.\ Hauert, A.\ Rossoll, and A.\ Mortensen.
\newblock {Fracture of high volume fraction ceramic particle reinforced
  aluminium under multiaxial stress}.
\newblock \emph{Acta Mater.}, 58\penalty0 (11):\penalty0 3895--3907, 2010.
\newblock \doi{10.1016/j.actamat.2010.03.037}.

\bibitem[Lou et~al.(2012)Lou, Huh, Lim, and Pack]{Lou2012a}
Y.\ Lou, H.\ Huh, S.\ Lim, and K.\ Pack.
\newblock {New ductile fracture criterion for prediction of fracture forming
  limit diagrams of sheet metals}.
\newblock \emph{Int.\ J.\ Solids Struct.}, 49\penalty0 (25):\penalty0
  3605--3615, 2012.
\newblock \doi{10.1016/j.ijsolstr.2012.02.016}.

\bibitem[Requena et~al.(2014)Requena, Maire, Leguen, and
  Thuillier]{Requena2013}
G.\ Requena, E.\ Maire, C.\ Leguen, and S.\ Thuillier.
\newblock {Separation of nucleation and growth of voids during tensile
  deformation of a dual phase steel using synchrotron microtomography}.
\newblock \emph{Mater.\ Sci.\ Eng.\ A}, 589:\penalty0 242--251, 2014.
\newblock \doi{10.1016/j.msea.2013.09.084}.

\bibitem[Samei et~al.(2016)Samei, Green, Cheng, and {de Carvalho
  Lima}]{Samei2016}
J.\ Samei, D.E.\ Green, J.\ Cheng, and M.S.\ {de Carvalho Lima}.
\newblock {Influence of strain path on nucleation and growth of voids in dual
  phase steel sheets}.
\newblock \emph{Mater.\ Des.}, 92:\penalty0 1028--1037, 2016.
\newblock \doi{10.1016/j.matdes.2015.12.103}.

\bibitem[Barsoum and Faleskog(2007)]{Barsoum2007}
I.\ Barsoum and J.\ Faleskog.
\newblock {Rupture mechanisms in combined tension and shear-Experiments}.
\newblock \emph{Int.\ J.\ Solids Struct.}, 44\penalty0 (6):\penalty0
  1768--1786, 2007.
\newblock \doi{10.1016/j.ijsolstr.2006.09.031}.

\bibitem[de~Geus et~al.(2016{\natexlab{a}})de~Geus, Cottura, Appolaire,
  Peerlings, and Geers]{DeGeus2016a}
T.W.J.\ de~Geus, M.\ Cottura, B.\ Appolaire, R.H.J.\ Peerlings, and M.G.D.\
  Geers.
\newblock {Fracture initiation in multi-phase materials: A systematic
  three-dimensional approach using a FFT-based solver}.
\newblock \emph{Mech.\ Mater.}, 97:\penalty0 199--211, 2016{\natexlab{a}}.
\newblock \doi{10.1016/j.mechmat.2016.02.006}.
\newblock \eprint{1604.03817}.

\bibitem[Cai et~al.(1985)Cai, Feng, and Owen]{Cai1985}
X.-L.\ Cai, J.\ Feng, and W.S.\ Owen.
\newblock {The dependence of some tensile and fatigue properties of a
  dual-phase steel on its microstructure}.
\newblock \emph{Metall.\ Trans.\ A}, 16\penalty0 (8):\penalty0 1405--1415,
  1985.
\newblock \doi{10.1007/BF02658673}.

\bibitem[Calcagnotto et~al.(2010)Calcagnotto, Ponge, and
  Raabe]{Calcagnotto2010}
M.\ Calcagnotto, D.\ Ponge, and D.\ Raabe.
\newblock {Effect of grain refinement to 1$\mu$m on strength and toughness of
  dual-phase steels}.
\newblock \emph{Mater.\ Sci.\ Eng.\ A}, 527\penalty0 (29-30):\penalty0
  7832--7840, 2010.
\newblock \doi{10.1016/j.msea.2010.08.062}.

\bibitem[Maresca et~al.(2014)Maresca, Kouznetsova, and Geers]{Maresca2014}
F.\ Maresca, V.G.\ Kouznetsova, and M.G.D.\ Geers.
\newblock {Subgrain lath martensite mechanics: A numerical-experimental
  analysis}.
\newblock \emph{J.\ Mech.\ Phys.\ Solids}, 73:\penalty0 69--83, 2014.
\newblock \doi{10.1016/j.jmps.2014.09.002}.

\bibitem[Steinbrunner et~al.(1988)Steinbrunner, Matlock, and
  Krauss]{Steinbrunner1988}
D.L.\ Steinbrunner, D.K.\ Matlock, and G.\ Krauss.
\newblock {Void formation during tensile testing of dual phase steels}.
\newblock \emph{Metall.\ Trans.\ A}, 19\penalty0 (3):\penalty0 579--589, 1988.
\newblock \doi{10.1007/BF02649272}.

\bibitem[Ghadbeigi et~al.(2010)Ghadbeigi, Pinna, Celotto, and
  Yates]{Ghadbeigi2010}
H.\ Ghadbeigi, C.\ Pinna, S.\ Celotto, and J.R.\ Yates.
\newblock {Local plastic strain evolution in a high strength dual-phase steel}.
\newblock \emph{Mater.\ Sci.\ Eng.\ A}, 527\penalty0 (18-19):\penalty0
  5026--5032, 2010.
\newblock \doi{10.1016/j.msea.2010.04.052}.

\bibitem[de~Geus et~al.(2016{\natexlab{b}})de~Geus, Du, Hoefnagels, Peerlings,
  and Geers]{DeGeus2016}
T.W.J.\ de~Geus, C.\ Du, J.P.M.\ Hoefnagels, R.H.J.\ Peerlings, and M.G.D.\
  Geers.
\newblock {Systematic and objective identification of the microstructure around
  damage directly from images}.
\newblock \emph{Scr.\ Mater.}, 113:\penalty0 101--105, 2016{\natexlab{b}}.
\newblock \doi{10.1016/j.scriptamat.2015.10.007}.
\newblock \eprint{1604.03814}.

\bibitem[Manigandan et~al.(2012)Manigandan, Srivatsan, and
  Quick]{Manigandan2012}
K.\ Manigandan, T.S.\ Srivatsan, and T.\ Quick.
\newblock {Influence of silicon carbide particulates on tensile fracture
  behavior of an aluminum alloy}.
\newblock \emph{Mater.\ Sci.\ Eng.\ A}, 534:\penalty0 711--715, 2012.
\newblock \doi{10.1016/j.msea.2011.11.081}.

\bibitem[Llorca and Segurado(2004)]{LLorca2004}
J.\ Llorca and J.\ Segurado.
\newblock {Three-dimensional multiparticle cell simulations of deformation and
  damage in sphere-reinforced composites}.
\newblock \emph{Mater.\ Sci.\ Eng.\ A}, 365\penalty0 (1-2):\penalty0 267--274,
  2004.
\newblock \doi{10.1016/j.msea.2003.09.035}.

\end{thebibliography}

\end{document}